%% file: fa_paper_nmf.tex
\documentclass[a4paper, top=2.5cm, bottom=2.5cm]{article} 
\usepackage[a4paper, top=3cm, bottom=2cm, left=2.5cm, right=2.5cm, marginparwidth=2cm]{geometry}

\usepackage[table, dvipsnames]{xcolor}
\usepackage{soul}
\usepackage[english]{babel}
\usepackage[utf8x]{inputenc}
\usepackage[T1]{fontenc}

\usepackage{amsmath}
\usepackage{setspace}
\usepackage{graphicx}
\usepackage[colorinlistoftodos]{todonotes}
\usepackage[colorlinks=true, allcolors=blue]{hyperref}
\usepackage{amssymb}
\usepackage{multirow}
\usepackage{dsfont}
\usepackage{amsthm}
\usepackage{siunitx}

\usepackage{stmaryrd}

\usepackage{subcaption}

\usepackage{natbib,har2nat}
\newcommand{\quantinar}{\raisebox{-1pt}{\protect \includegraphics[scale=0.04]{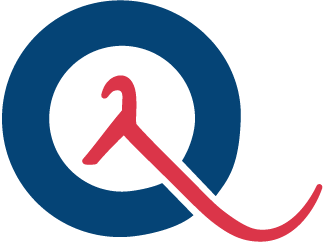}}\,}
\newcommand{\quantletNMFRB}{\href{https://github.com/QuantLet/EmbeddingPortfolio/tree/main/NMFRB}{\protect \includegraphics[scale=0.05]{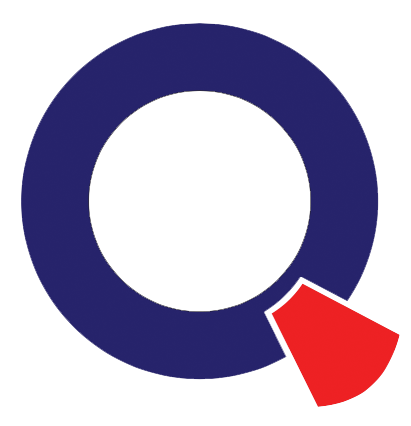} NMFRB}}

\providecommand{\keywords}[1]
{
	\small	
	\textbf{\textit{Keywords:}} #1
}
\providecommand{\classcodes}[1]
{
	\small	
	\textbf{\textit{JEL classification:}} #1
}

\title{Risk budget portfolios with convex Non-negative Matrix Factorization}

\author{
	Bruno Spilak \footnote{corresponding author, IRTG 1792, School of Business and Economics, Humboldt-Universit{\"a}t zu Berlin, Doroheenstr. 1, 10117 Berlin, Germany. Email: bruno.spilak@hu-berlin.de},
	Wolfgang Karl Härdle\footnote{IRTG 1792, School of Business and Economics, Humboldt-Universit{\"a}t zu Berlin, Dorotheenstr. 1, 10117  Berlin,  Germany; Humboldt-Universit{\"a}t zu Berlin, Blockchain Research Center (BRC), Berlin, Germany; Institute for Digital Assets (IDA), Bucharest University of Economic Studies, Bucharest, Romania; Sim Kee Boon Institute, Singapore Management University, Singapore; Asian Competitiveness Institute (ACI), NUS, Singapore; Dept. Information Management and Finance, Yushan Scholar National Yang-Ming Chiao Tung University, Hsinchu, Taiwan, ROC; Dept. Mathematics and Physics, Charles University, Prague, Czech Republic; the European Cooperation in Science \& Technology COST Action [CA19130]; Grant CAS: XDA 23020303 and DFG IRTG 1792 gratefully acknowledged}
}

\begin{document}
\maketitle

\begin{abstract}

We propose a portfolio allocation method based on risk factor budgeting using convex Nonnegative Matrix Factorization (NMF). Unlike classical factor analysis, PCA, or ICA, NMF ensures positive factor loadings to obtain interpretable long-only portfolios. As the NMF factors represent separate sources of risk, they have a quasi-diagonal correlation matrix, promoting diversified portfolio allocations. We evaluate our method in the context of volatility targeting on two long-only global portfolios of cryptocurrencies and traditional assets. Our method outperforms classical portfolio allocations regarding diversification and presents a better risk profile than hierarchical risk parity (HRP). We assess the robustness of our findings using Monte Carlo simulation.

\end{abstract}

\keywords{Portfolio management, factor model, non-negative matrix factorization, risk budget, risk parity}

\classcodes{C10, C14, C21, C22, C45, C58, G10, G11}

\newpage

\input{introduction_nmf}
\input{methodology_nmf}
\input{oos_results_nmf}
\input{mc_sim_nmf}
\input{conclusion_nmf}

\bibliography{../bibliography}

\newpage
\input{appendix_nmf}


\end{document}

%% file: introduction_nmf.tex
\section{Introduction}
\label{sec:intro}

Portfolio selection is at the heart of robust wealth management and is one of the main tasks of "quants". Its goal is to find an allocation of capital across various assets that maximizes the portfolio returns within an acceptable level of risk. Modern Portfolio Theory suggests that diversification reduces the portfolio risk, measured in terms of variance, and benefits from a low correlation of the assets. The literature around this problem is more than vast, however, there are commonly two main approaches: a "model-free" and "model-based" approach \citep{Pastor:2000}.

The first one assumes functional forms of the stochastics of asset returns; an approach taken by \citet{Markowitz:1952} that led him to critical line algorithms (CLA), which are based on sample estimates of the mean and covariance of the asset returns. Practitioners have recognized the limits of this approach. A major caveat is that small deviations in the estimate of the returns mean will cause CLA to produce very different portfolios \citep{Michaud:2008}. On top, the original mean-variance method requires the inversion of the covariance matrix, which is prone to large errors when the covariance matrix is numerically ill-conditioned \citep{BaileyPrado:2012}. To obtain stabler portfolio allocation, many alternatives have been proposed, see \citet{Kolm:2014, black1992global} among others. After the 2008 crisis, a lot of attention has been given to the Risk Contribution Portfolio (RCP), also known as Risk Budget Portfolio (RBP) or risk parity. The RCP allocates risk among the assets instead of capital following a risk budget rule defined by the portfolio manager \cite{Maillard60, Roncalli:2013}. The goal of RCP is to find the capital allocation where the marginal risk contribution of each asset corresponds to a predefined budget. The Risk Parity (RP) or Equal Risk Contribution (ERC) portfolios correspond to the simplest risk budget rule where the risk contribution is the same for all assets in the portfolio. When the assets are perfectly uncorrelated, \citet{HKPZ:2021} show that the RP corresponds to an Inverse Variance (IV) allocation at the asset level and is optimal for Markowitz's type investors. Thus, finding uncorrelated sources of risk provides a solution to the portfolio allocation problem. A traditional approach for global portfolios managed by risk parity funds allocates the risk equally to different asset classes like stocks, bonds, and commodities, assuming the asset classes are uncorrelated. In order to avoid this naive budget rule, the Hierarchical Risk Parity (HRP) approach \cite{Prado:2016} introduces a hierarchy in the correlation matrix to build a diversified portfolio with a risk parity allocation using tree clustering. On top of being purely data-driven, an advantage of HRP is that it does not require the inversion of the covariance matrix to compute the portfolio weights, as in the \href{https://www.quantinar.com/course/135/robustifying-markowitz}{\quantinar Robust Markowitz} method, making them more stable. However, as has been underlined by \citet{Raffinot:2017}, besides suffering from certain disadvantages of minimum spanning tree (MST) and single linkage with chaining, HRP does not leverage the structure of the dendrogram obtained from the tree clustering. Raffinot overcomes that issue by directly using the dendrogram shape for the portfolio allocation in his Hierarchical Clustering Asset Allocation (HCAA) method.  

Nevertheless, model-free approaches rely mostly on the study of the returns covariance matrix structures and ignore the potential usefulness of "model-based" approaches, such as factor models. 

In this second approach, the investor uses his prior knowledge to build a linear model for the returns as follows: $x_t = Wz_t + \varepsilon_t$, where $t \in [1, T]$, $x_t \in \mathbb{R}^{d,1}$ is a vector of returns, $W \in \mathbb{R}^{d,p}$ the unknown factor loading matrix corresponding to the factors $z_t \in \mathbb{R}^{p,1}$ for $p \ll d$ and $\varepsilon_t$ correspond to the idiosyncratic errors. Such a linear model tells us that the asset returns can be explained by a few specific variables, which are either defined by prior established knowledge about the empirical behavior of average returns \citep{Kelly:2018}, as in the Fama and French model \citep{FAMA1993}; or they are treated as latent variables and directly estimated from the data as a linear combination of the asset returns, using factors analysis,\href{https://www.quantinar.com/course/544/mva-principal-components}{\quantinar Principal Component Analysis} (PCA) \cite{partovi2004principal} on the covariance matrix of returns or \href{https://www.quantinar.com/course/243/deda-independent-component-analysis}{\quantinar Independent Component Analysis} (ICA) \cite{lassance2022optimal}. This method based on statistical factors has the main advantage of requiring no ex-ante knowledge of the structure of average returns and explains the correlation between assets, which must be considered to build a diversified portfolio. Indeed, the RP portfolio is optimally diversified in terms of risk only if the asset returns are uncorrelated. To overcome this, researchers have turned to the Factor Risk Parity portfolio (FRP) which assumes a linear factor model for the returns in order to spread the portfolio risk equally among the uncorrelated factors, see e.g. \cite{meucci2009managing, Deguest2013RiskPA, lohre2014diversifying, roncalli2016risk, lassance2022optimal} and \cite{ kelly2023principal} in the context of asset pricing. By its very nature, PCA is used to find uncorrelated factors. However, PCA suffers from a few well-known limitations explained by \citet{Meucci2015RiskBA} in the context of portfolio optimization. The factors are not robust, especially those related to the lowest eigenvalues, since they maximize the variance which is sensitive to outliers and they are not unique. To remove those factors, \citet{meucci2009managing} proposed to include them inside the set of constraints $C$ so that their exposure is 0. This is an approach taken by \citet{roncalli2016risk}. Yet an other solution is to use ICA as factorization method as in \citet{lassance2022optimal} which may be more robust than PCA since the optimization in ICA is related to the L1 and not the L2-norm as for PCA \cite{martin2016link}. However, both ICA and PCA can be difficult to interpret in a long-only setting because their loadings are not necessarily positive. A fortiori ICA factors cannot be interpreted with respect to long-short positions as it is invariant to a change of sign. 

In this paper, we contribute to the literature on correlation matrix clustering and factor models by investigating non-negative matrix factorization (NMF) technique to build a new FRP allocation. We restrict our analysis to long-only portfolios as we are interested in building a global portfolio including cryptocurrencies which, at times of writing, are still very difficult to short in practice. Thanks to the design of the factor model, we build a long-only portfolio allocation from the factor loading matrix that is intuitive since, in the linear case, the factors are interpretable as cluster centroids and long-only portfolios, and guarantees diversification at the factor and individual asset levels since the factors are low-correlated out-of-sample. Instead of using the traditional PCA or ICA, the factor model is designed in a specific fashion by adding two constraints to the traditional one: the factor loadings must be positive $W\in \mathbb{R}_{+}^{d}$, which allows us to find assets that are positively correlated, and the factors lie within the column space of the asset returns: $z_t = Hx_t$ where $H \in \mathbb{R}_{+}^{p,d}$, so we could interpret $z_t$ as portfolios capturing the notion of centroids. This approach corresponds to the convex Non-negative Matrix Factorization (NMF) of the returns, a technique developed by \citep{Ding:2010}. Convex NMF extends the classical NMF method \cite{Lee1999} to mixed signs data, such as financial returns. NMF has great success in various applications such as text mining \citep{doi:10.1137/1.9781611972740.45} or pattern recognition \citep{990477}, but its application in finance is mostly restricted to price data analysis \citep{Drakakis:2008}. \citet{Frein:2008} proposed to use a sparse variant of NMF for clustering stocks into a latent trend-based domain, as opposed to the traditional sector-based domain to build better-diversified portfolios. Similarly, we propose to use the inherent sparsity and clustering ability of the convex NMF approach \cite{Ding:2005, Ding:2010} to easily cluster the assets which are positively correlated to the underlying factor by exploiting the structure of the covariance matrix of the asset returns. Even if NMF is not unique in general (see e.g. \cite{NIPS2003_1843e35d, Laurberg2008TheoremsOP, 6630130}), as PCA, Convex NMF is naturally sparse and enforces the uniqueness of the factors. Its enhanced interpretability over PCA and ICA makes it particularly appealing for the long-only portfolio allocation problem.

Our proposed portfolio allocation is evaluated in two universes. The first global portfolio consists of bonds, stock indices, commodities, Forex pairs, and cryptocurrencies, which have been gaining strong interest from traditional investors in the last two years \citep{doi:10.1080/14697688.2021.1880023}, The second universe is the same as in \citet{Raffinot:2017} and consists of bonds at various tenors, stock, and commodity indices on a 30-year long time frame. We backtest our portfolio using a dynamic volatility target strategy, a traditional approach for risk parity portfolios. Our Python implementation and results are available on \href{https://github.com/QuantLet/EmbeddingPortfolio}{quantlet.de} (\includegraphics[scale=0.05]{quantlet.png} \href{https://github.com/QuantLet/EmbeddingPortfolio/}{EmbeddingPortfolio}) for reproducible and transparent research.

Our findings show that on the two universes, our method is successful at finding robust lowly correlated sources of risk out-of-sample across our test sets and enhancing the diversification between and within them. It outperforms HRP method in terms of risk-adjusted returns and naive strategies in some specific cases, an important incentive for portfolio managers to integrate machine learning into their portfolio allocation methodologies universe. The structure of the paper is as follows: first, the model and portfolio allocation methodology are described. In the second part, we present the evaluation design, then we present the main empirical findings. Finally, we perform some robustness test using Monte Carlo simulation, comparable to the one performed in \citet{Prado:2016}.

%% file: methodology_nmf.tex
\section{Non-negative matrix factorization for portfolio allocation}
\subsection{(Convex) Non-negative matrix factorization}\label{sec:convex_NMF}

One main aspect of portfolio management is diversification, which is defined as spreading investments into a variety of assets that have different risk exposures to increase the return while reducing the risk of the portfolio. One classical measure of linear co-dependence is correlation. Thus, diversification aims at finding assets that present low correlation. A simple approach is to build a global portfolio with various asset classes, for example, bonds, stocks and cryptocurrencies, since they are driven by different fundamentals. However, those three asset classes have different volatility levels. In fact, cryptocurrencies have very large volatility compared to bonds or even stocks: BTC has 73\% annualized vol in the period Aug 2015 to Nov 2021, 18\% for SP500 and 6\% for the US 10Y bond, which is 4 and 12 times larger than SP500 and US 10Y bond, respectively. Using a naive risk parity allocation would reduce the exposure to the cryptocurrency market to 0. 

Moreover, instead of relying on naive segmentation of the portfolio basket based on asset classes, we aim at finding one that achieves diversification without prior economic knowledge. The goal is then to learn how to map, for any $t,\ 1\leq t \leq T$, the $d$ asset returns $r_t = (r_{t,1}, \ldots, r_{t,d})  \in \mathbb{R}^{d}$ into $p \leq d$ separate factors $f_t = (f_{t,1}, \ldots, f_{t,p})\in \mathbb{R}^p$  with low inter-correlation and large intra-correlation. That way each asset must have high exposure to a single factor $k$ which can be interpreted as the cluster centroid of the assets that are projected on itself. Semi and convex non-negative matrix factorization (NMF) achieves that goal.

Let us first briefly present the simple NMF approach from \cite{Lee1999}. Describing the problem in a matrix factorization framework with $X \in \mathbb{R}_+^{T, d}$ a data matrix, the goal of NMF, as PCA, is to find a factor matrix $Z  \in \mathbb{R}_+^{T, p}$ and associated factor loading matrix $W  \in \mathbb{R}_+^{p, d}$ , of size $T \times p$ and $p \times d$ respectively, such that $X = ZW^{\top} + \mathcal{E}$ where $\mathcal{E} \in \mathbb{R}^{T, d}$ is the matrix of residuals. The rank $p$ of the factorization is a hyperparameter and must be chosen so that $p \ll d$. We will explain how to select it in section \ref{sec:hyperparameters}.

The difference between PCA and NMF lies in the constraints imposed on the factorization. Indeed, PCA forces the rows of $W$ to be orthonormal and the columns of $Z$, the principal components, to be orthogonal, while NMF applies to data matrix $X$ with positive entries only and does not allow negative ones in the matrix factors $W$ and $Z$. In contrast to PCA, in the linear combination of the factors, only additive operations are allowed, which makes NMF more interpretable than PCA and intuitively compatible with the notion of parts-based representation.

It is impractical to apply NMF to mixed-sign data such as financial returns, thus \citet{Ding:2010} proposed a semi-NMF method in which $Z$ is unconstrained while $W$ is restricted to be nonnegative. On top, convex-NMF constrains the vectors defining $Z$ to lie within a convex column space of $X$ so that the columns of $Z$, the factors $z_k$, can be interpreted as long-only portfolios, that is: $Z = XH$ with $H \in \mathbb{R}_+^{d, p}$. $W$ and $H$ can be estimated by minimizing the reconstruction error $||X - XHW^\top||_2$. This optimization problem is non-convex and a solution via multiplicative update is given in Algorithm 3.2 from \citet{Ding:2010}.

Convex NMF tends to produce very sparse matrices $H$ and $W$, which effectively allows for a parts-of-whole representation of the data. On top, \citet{Ding:2005, Ding:2010} show that, in practice, $W$ and $Z$ in convex NMF are naturally close to orthogonal, which implies that the factors in $Z$ are lowly correlated. In our context, convex NMF can then represent a portfolio of $d$ assets with $p$ lowly correlated synthetic assets. This can also be interpreted as a clustering result. If $W$ is indeed orthogonal ($W^\top W = I$), then NMF and convex NMF are relaxations of $K$-means clustering and \textit{soft} relaxations in the general case (theorem 5 and 3 from \citet{Ding:2005, Ding:2010} respectively). Thus, NMF is expected to find clusters in the asset universe with centroids encoded by $Z$ similar to $K$-means and cluster assignment probabilities encoded by $W$. $W$ can be directly used for cluster assignment as orthogonality and nonnegativity together imply that each row of $W$ has only one non-negative element, which allows to interpret $W$ as encoding cluster indicators which is not the case for PCA or ICA. Let $C_k$ denote the cluster that includes all asset returns projected on factor $z_k$, it is defined as $C_k = \{i\}_{1\leq i \leq d,\ \max(W_{i:}) = k}$.

\begin{figure}[!ht]
	\centering
	\begin{subfigure}[b]{0.32\textwidth}
		\centering
		\includegraphics[width=\textwidth]{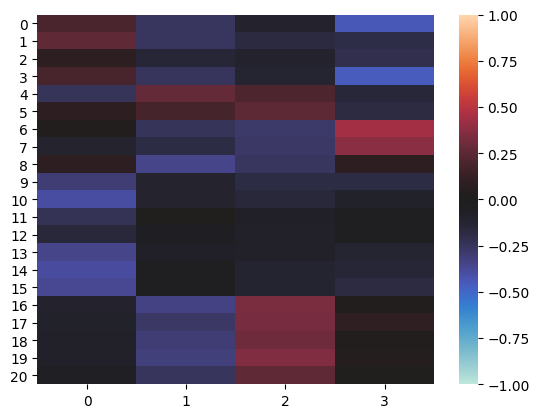}
		\caption{PCA}			
		\label{fig:ex_loadings_pca}
	\end{subfigure}
	\hfill
	\begin{subfigure}[b]{0.32\textwidth}
		\centering
		\includegraphics[width=\textwidth]{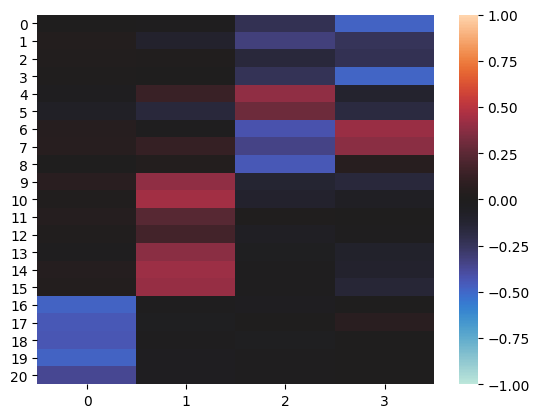}
		\caption{ICA}			
		\label{fig:ex_loadings_ica}
	\end{subfigure}
	\hfill
	\begin{subfigure}[b]{0.32\textwidth}
		\centering
		\includegraphics[width=\textwidth]{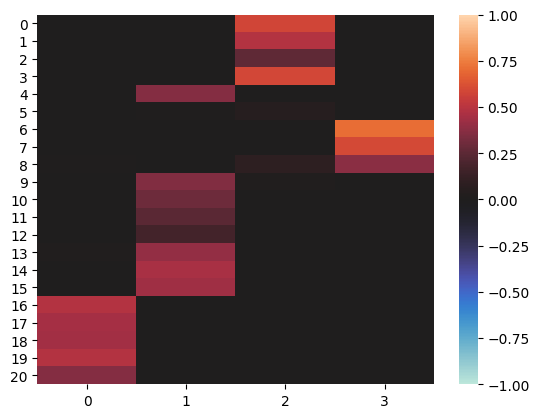}
		\caption{NMF}			
		\label{fig:ex_loadings_nmf}
	\end{subfigure}
	\caption[]{Example of matrix factor loadings $W$ for PCA, ICA and convex-NMF}
	\label{fig:ex_loadings}
\end{figure}

For illustration, we show in Figure \ref{fig:ex_loadings}, the matrix of factor loadings on a dataset with 21 assets for PCA, ICA and convex-NMF. The sparsity of $W_{\operatorname{NMF}}$ and the separation of the factors in 4 distinct long-only portfolios are obvious. Ignoring the sign of the loadings, the ICA decomposition retrieves the same cluster centroids as NMF for the first two factors. Finally, the PCA factors are difficult to interpret as they combine long-short portfolios, in particular, we see that factor 1 and 2 have similar loadings except for the last 5 assets that have negative loadings in factor 1, but positive ones in factor 2.

An interesting note is whether to apply NMF on standardized data or raw data. In terms of reconstruction error, as NMF is scaling-invariant, NMF will perform best with standardized data if the raw data is dominated by a few variables that have a large scale. Indeed, if the data have different scales, which is the case in our context of a global portfolio that includes cryptocurrencies, the factors may be biased towards the assets with larger variance. On top, standardizing the data can improve the interpretability of the factors, since it allows the comparison of the contributions of different assets to each factor directly. Finally, performing NMF on standardized data facilitates the interpretation of the factors as cluster centroids as we explain in the next paragraph.

We can easily show that clustering assets based on the pairwise Euclidean distance of their standardized returns $d_E(r_i^*, r_j*)$, where $r_i^*= \Bigl( \sigma^{-1}(r_{t} - \mu)\Bigr)_{i}$ with $\mu = (\mu_i)_{1\leq i \leq d}$ and $\sigma = \operatorname{diag}(\sigma_i)_{1 \leq i \leq d}$ are respectively the sample mean vector and the diagonal matrix of the individual standard deviation of the $d$ assets, has a direct interpretation in terms of correlation. Indeed, $d_E(r_i^*, r_j^*)$ is related to the Pearson correlation coefficient, $\rho(r_i, r_j)$ with the following equality: $d_E(r_i^*, r_j^*) = \sqrt{2T\{1 - \rho(r_i, r_j)\}} \propto \sqrt{\{1 - \rho(r_i, r_j)\}}$. Thus, performing a clustering using the Euclidean distance on the standardized returns matrix $X^*$ is equivalent to a clustering using the distance $d_{\rho}(r_i, r_j) = \sqrt{\{1 - \rho(r_i, r_j)\}/2}$ on the raw matrix $X$. This distance is maximal for perfectly negatively correlated assets and minimal for uncorrelated ones and it is used in the tree-clustering of the HRP portfolio from \citet{Prado:2016}. As convex NMF is a (soft) relaxation of K-means, we can expect it to cluster assets that are highly correlated together and separate assets that are negatively or lowly-correlated.  Thus, we perform convex NMF of the standardized returns matrix, we have: $X^*= ZW^{\top} + \mathcal{E}$. 

The NMF decomposition is however not unique since $ZW^{\top} = ZDD^{-1}W^{\top} = Z'W'^{\top}$ where $D$ is a nonnegative matrix such that $D^{-1}$ is nonnegative (see the literature about identifiability criterion for NMF decomposition, e.g. \cite{NIPS2003_1843e35d, Laurberg2008TheoremsOP, fu2018identifiability}). Nevertheless, neglecting unavoidable permutation and scaling indeterminacy, a unique NMF solution can be obtained. We need to enforce at least one of the following techniques: normalizing the input data $X$ so it has zero-mean, the columns of $W$ to unit norm and imposing sparsity constraints to $Z$ or $W$ \cite{doi:https://doi.org/10.1002/9780470747278.ch1}. We already know that $X^*$ has zero mean and $W$ is sparse. In order to enforce the columns of $W$ to have unit norm, we simply replace $W$ by $WD^{-1}$ in the multiplicative updates of Algorithm 3.2 from \citet{Ding:2010} with $D = \operatorname{diag}(\lVert W_{:,k} \rVert_2)_{1\leq k \leq p}$. The diagonal entries of $D$ represent the variance explained by the factors $z_k$ of the standardized data. We now consider that $W$ columns have unit norm. We explain why this constraint is useful for the portfolio allocation in Section \ref{sec:nmfrb}.

We can retrieve the linear factor model for the raw returns vector $r_t$ from the latent factors $z_t^*$ as it follows:
\begin{align}
	r_t^* &= Wz_t + \varepsilon_t \nonumber \\
	\sigma^{-1} (r_t - \mu) &= Wz_t  + \varepsilon_t \nonumber \\
	r_t &= \sigma Wz_t + \sigma \varepsilon_t + \mu \nonumber \\
	r_t &= \widetilde{W}z_t  + \mu + \eta_t \label{eq:nmf_model}
\end{align}

where:
\begin{align}
	\widetilde{W} &= \sigma W \label{eq:W} \\
    \eta_t & = \sigma \varepsilon_t\\
    \Vert w_{:k} \rVert_2 &= 1,\ \forall 1 \leq k \leq p \label{eq:cons_unit}
\end{align}
Then, $W$ is the normalized factor loadings matrix and the factor loadings $w_{ij}$ represent the correlation between the original assets returns $i$ and the factors $j$ and the squared loadings $w_{ij}^2$ indicates what percentage of the variance in the original asset $i$ is explained the factor $j$. 

As in PCA, we can represent a portfolio allocation $a$ in the assets space with an allocation $\tilde{a}$ in terms of the NMF lowly-correlated portfolios $z_t$ with $\widetilde{a}_t = \widetilde{W}^\top a_t$. We have: $R_t(a) = a^{\top}r_t = a^{\top}Wz_t + a^{\top}\eta_t = \widetilde{a}_tz_t + e_t$. However, as $W$ is not invertible, the expression of of $a_t$ in terms of the NMF factors allocation is more complex than for PCA and is given by \citet{roncalli2016risk}. Let’s note $V$ = $\widetilde{W}^T$ and $V^+$ the Moore-Penrose inverse of $V$. We have therefore $a = V^+\widetilde{a} + e$ where $e = (I_d - V^+V)a \in \mathbb{R}^{d, 1}$ is in the kernel of $\widetilde{V}$. We also have $\Sigma = W \Omega W^\top$ where $\Omega$ the covariance matrix of the factor returns $z_t$.

\subsubsection{Hyperparameters}\label{sec:hyperparameters}

NMF has one hyperparameter, $p$ the number of latent factors. As $p$ approaches $d$, the reconstruction error of NMF will decrease, nevertheless, any meaningful information from the factors is lost and a model with a large $p$ overfits. This is a classical model selection problem and there are multiple ways to select $p$. In a parametric setting, a good criterion is the Akaike Information Criterion (AIC) \cite{akaike1970statistical} that uses the maximum likelihood. However, we do not make any assumption on the distribution of the residuals, $\varepsilon$, in our model, and the likelihood is not available. We must estimate the density of the residuals, for example, with a Kernel Density Estimator (KDE), to use the AIC in a non-parametric setting. This approach is taken by \citet{trimborn2018crix} to estimate AIC and select the number of constituents in the cryptocurrency index, CRIX. However, in the CRIX method, $\varepsilon$ has dimension one whereas, in our case, it has dimension $d$ which can be very large for a portfolio. This is problematic because estimating the KDE of a high-dimensional density is challenging due to the curse of dimensionality and the choice of bandwidth \citep{Haerdle2004}. An alternative, widely adopted in Machine Learning (ML) is cross-validation, but it requires a vast amount of data, in particular for time series, which are not available in our case.

As NMF is a relaxation of $K$-means we propose to use tools from the clustering literature (see \href{https://www.quantinar.com/course/102/all-about-clustering}{\quantinar All about clustering!}) about the optimal number of clusters, e.g. the gap statistic \cite{tibshirani2001estimating} or the average of the silhouette coefficient \cite{ROUSSEEUW198753}. The gap statistic is computationally intensive to compute as it requires multiple Monte Carlo estimates drawn from a reference distribution, thus, we propose to compare the average silhouette coefficient. The silhouette coefficient for a sample $i$ with $p$ clusters is defined as:
$$
s_i(p) = \frac{b - a}{\max(a, b)} 
$$
where $a$ is the average distance between sample $i$ and all other points in the same cluster and $b$ is the average distance between sample $i$ and all other points in the next nearest cluster that does not include $i$.
For all $i$, $-1 \leq s_i(p) \leq 1$, if $s_i$ is close to 1 this implies that the intra-cluster distance of sample $i$ with other assets in the same clusters is much smaller than the smallest nearest-cluster distance for each sample. Therefore, we can say that $i$ is well-clustered.
The silhouette score computes the average silhouette coefficient for all samples:
\begin{equation}\label{eq:silhouette_score}
	\mathcal{S}(p) = 1/d \sum_{i=1}^d s_i(p)
\end{equation}
and we could obtain the optimal number of cluster $p^*$ by maximizing the silhouette score for the candidates $p$: $p^* = \max_{2 \leq p \leq d - 1} \mathcal{S}(p)$. To compute the silhouette coefficients, we use the Euclidean distance which, as we have seen in the previous section, is equivalent to the distance $d_{\rho}$. Thus, the silhouette coefficient has an interpretation in terms of correlation: a high silhouette coefficient for sample $i$ implies that it has a much larger pairwise correlation with the assets in its cluster than the largest correlation with another asset $j$ in the nearest cluster. Maximizing the silhouette score should then give a clustering with a large intra-cluster correlation and low inter-cluster correlation.

As we know that the correlation structure is not stable in time, we can expect the optimal number of clusters to be highly dependent on the historical interval the model is trained on. Since a good model needs to learn general patterns that do not depend on a specific observed path, but account for the non-stationarity in the returns, new samples are generated with the block-bootstrap technique. Indeed, block-bootstrap helps to preserve stylized facts of asset returns, such as serial correlation
and heteroscedasticity. Following \cite{Jaeger:2021}, we construct a new return time series, with the same length as the original, by sampling blocks with replacement and a fixed length (60 days), but a random starting point. We repeat this operation $B$ times, for example, $B=1000$, and for each time series, we estimate $\mathcal{S}(p)^b$ with various $p$.  For each $p$, we obtain the silhouette score with the average bootstrap estimates $\mathcal{S}(p) = 1/B \sum_b \mathcal{S}(p)^b$ and its standard deviation, $\sigma(p) = \sqrt{1/B \sum_b \left( \mathcal{S}(p)^b - \mathcal{S}(p)\right)^2}$. We select $p^*$ as the extrema of the silhouette score decision curve $p \rightarrow \mathcal{S}(p)$ which maximizes the bootstrap silhouette Sharpe ratio: $\mathcal{S}(p) / \sigma(p)$. That way, $p^*$ is a robust choice since it has a low bootstrap variance and clusters the assets well since it has a large silhouette score.

Finally, because of the non-stationarity of the returns, we can also expect the optimal number of clusters to be variable in time. For example, in periods of crisis, assets from different classes tend to be more correlated which reduces the number of clusters. To obtain a dynamic number of clusters, we recompute $p^*$ at each monthly rebalancing period. We comment the dynamics of $p^*$ in Section \ref{sec:empirical_results} Figure \ref{fig:n_factors}.

Let us now present the portfolio allocation.

\subsection{NMF for diversified portfolio allocation}
There are multiple ways to allocate portfolios that diversify risk based on the NMF model, see \citet{roncalli2016risk} for an extensive review of risk diversification methods with risk factors. In this article, our goal is to obtain a portfolio allocation $a$ which is diversified at both the NMF factors level, corresponding to synthetic long-only portfolio or asset classes, and the original assets. We focus on the volatility risk measure since it is the most widely used measure in modern portfolio theory. In contrast to other approaches like those mentioned in \cite{meucci2009managing, lassance2022optimal, lohre2014diversifying}, our proposed allocation does not require additional optimization beyond the training of the NMF model. We achieve our goal by making simple and reasonable assumptions on the NMF factors, as in Hierarchical clustering Risk Parity (HRP) \cite{Prado:2016} and Hierarchical clustering Asset Allocation (HCAA) \cite{Raffinot:2017}. This approach significantly simplifies the implementation of our method. Before presenting our approach, we will provide a brief overview of the HRP and HCAA portfolio allocations.

The HRP introduces a hierarchy in the correlation matrix to build a diversified portfolio. An advantage of HRP is that it does not require the inversion of the covariance matrix to compute asset weights. The algorithm consists of three steps. First, by using tree-clustering, the algorithm finds a linkage matrix based on the correlation to group assets together in clusters. Then, one performs quasi-diagonalization, also called matrix seriation, of the correlation matrix to place assets that are highly correlated together along the first diagonal and the ones that are not correlated far apart from each other to obtain a quasi-diagonal correlation matrix. When the covariance matrix is diagonal, the optimal allocation, for a Markowitz-type investor, corresponds to inverse-variance weights (see Chapter 19 in \citet{mva:2019}). Thus, the final step of the algorithm uses a top-down recursive bisection to split allocations between adjacent clusters in inverse proportion to their aggregated variances. As it has been underlined by \citeauthor{Raffinot:2017}, on top of suffering from certain disadvantages of minimum spanning tree (MST) and single linkage with chaining, HRP does not leverage the structure of the dendrogram obtained from tree-clustering. Indeed, MST is only used to reorder assets, but the dendrogram is not used when building the allocation with bisection, only the number of assets matters. In his paper, \citeauthor{Raffinot:2017} overcomes that issue by directly using the dendrogram shape for the bisection in his HCAA method. The author find a diversified weighting by distributing capital equally to each cluster, so that many correlated assets receive the same capital as a single uncorrelated one. Then, within a cluster, an equal-weighted allocation is computed.

In this paper, we propose to leverage the learned structure from the NMF factor model to diversify risk among the factors.

\subsubsection{Convex NMF Risk Parity (NMFRP)}
First, we present the convex NMF Risk Parity (NMFRP) portfolio, which is a naive allocation based on a risk parity approach, like the HRP. Since the obtained factors present low pairwise correlations, we can assume that the associated cluster portfolios have a quasi-diagonal covariance matrix. Thus, we apply a simple inverse-variance allocation at the cluster level, because it is optimal for a Markowitz-type investor.

Let us denote $a^j = (a_i^j)_{i\in C_j}$ the intra-cluster weight vector where $a_i^j$ is the weight of asset $i$ within the cluster $C_j$. We can compute the variance of the associated portfolio as $\sigma_{fj}^2(a)= {a^j}^{\top}\Sigma_{j}a^j$ where $\Sigma_{j}$ corresponds to the covariance of cluster $\mathcal{C}_j$.  Let us denote $c_j$ as the inter-cluster weight, we set:
$$c_j = {\sigma_{fj}^2}(a) ^{-1} / \sum_{k=1}^p {\sigma_{fk}^2}(a)^{-1}$$

For the intra-cluster weights, we propose to use an Equal Risk Contribution (ERC) allocation. Since the assets are highly correlated within each cluster $C_j$, we can assume that the pairwise correlations are equal and elevated. \citet{Maillard60} shows that weights of an ERC portfolio do not depend on the correlations of the asset if they are equal ($\rho_{ij} = \rho$ for all $i,j$). In that case, the ERC portfolio is obtained with the inverse volatility allocation: 
$$a_i^j = \frac{\sigma_i^{-1}}{\sum_{k \in C_j} \sigma_k^{-1}}$$
where $\sigma_i$ is the volatility of the asset $i$. If on top we assume that the assets within each cluster have the same risk-adjusted expected return, which can be estimated with the Sharpe ratio ($\operatorname{SR} = \mu_i/\sigma_i$ for asset $i$), then the ERC allocation is optimal for a Markowitz type investor. The final weight for each asset is simply the intra-cluster weight rescaled by its associated inter-cluster weight, that is $a_i = a_i^j \times c_j$. 

In other words, the NMFRP allocation is simply the inverse-variance weight at the cluster level rescaled by the individual inverse-volatility weight within each cluster. This allocation reflects the disparities between asset classes more accurately than the simple inverse-variance allocation. Indeed, when we are dealing with a global portfolio with various asset classes, for example, bonds, stocks, cryptos, and Forex, the goal is to take advantage of the diversification effect from the relatively low correlation between asset classes. Nevertheless, they might have different volatility levels. Typically, Forex and bonds have low volatility while stocks have high and cryptos have extremely high volatility. By applying an inverse-variance allocation to this universe, cryptos would have close to 0 allocations, neutralizing any potential diversification effect from the crypto class. When using an allocation at the cluster level, the assets belonging to another cluster are disregarded when computing the individual weights. HRP and the proposed NMFRP achieve that goal.

\subsubsection{Convex NMF Risk Budget (NMFRB)}\label{sec:nmfrb}

The main issue with the previous method is that it does not leverage the learned factor structure from the model. As in HRP, NMF gives a clustered representation of the correlation structure between the assets, but with one main advantage.  Indeed, as stated in Section \ref{sec:convex_NMF}, the factor loadings ($w_{ij}$) tell which assets are well explained by the factor model in terms of correlation. Thus, the selection of assets that explain most of the correlation structures in the basket is clear. Those assets will have a higher loading: the highest being 1 and the lowest 0, due to unit norm and positive constraints on the loadings. In other words, convex NMF assigns a 0 weight to the assets that do not explain any correlation, if the number of factors $p$ is not too large. Economically, this behavior can be desirable, as having too many assets in a frequently rebalanced portfolio causes high transaction costs if the turnover is high. We propose therefore to take into consideration the actual factor loading values, $w_{ij}$, and to allocate more risk to the assets that are well explained by the model, that is, the ones that have a large exposure to the latent factors.

To do so, we use a Risk Budget (RB) allocation \cite{bruder2012managing, Roncalli:2013} within each cluster $C_j$ using the budget constraints $b_i = (w_{ij}^2)_{i\in C_j}$. This forms the $\operatorname{NMFRB}$ portfolio that allocates the risk to asset $i$ proportionally to the fraction of its variance explained by the factor $j$, expressed by $w_{ij}^2$, within each cluster $j$. It is clear that the risk budgets sum to one since each column of $W$ has a unit norm (see Equation \eqref{eq:cons_unit}). We could solve the classical optimization program for RB portfolios to find the final capital allocation within each cluster. However, we reduce the difficulty of the problem by assuming that the pairwise correlations of the assets within each cluster are equal to 1 since we know that within each factor portfolio, the correlation is elevated. Then, we know from \citet{bruder2012managing} that $a_i^j = b_i^j \sigma_i^{-1} / \sum_{k \in C_j} b_k^j \sigma_k^{-1} = w_{ij}^2 \sigma_i^{-1} / \sum_{k \in C_j} w_{kj}^2 \sigma_k^{-1}$. Thus, the asset weight within each cluster corresponds to the proportion of the variance explained by the associated factor scaled by the inverse of the asset's volatility. This method allocates more capital to the assets that are well explained by the model and have low volatility and the ones that are not explained will have a weight close to 0.

Finally, we know that the correlations at the cluster level are low, but there is no guarantee that they are 0 and they are probably not. Thus, an allocation in terms of ERC rather than in terms of inverse variance seems more appropriate. As for the intra-cluster weights, we can assume that the clusters have low and equal correlations, thus the ERC allocation is given in terms of inverse volatility:
$$c_j = \sigma_{fj}^{-1} / \sum_{k=1}^p \sigma_{fk}^{-1}$$
and the final capital allocation for each asset is given by $a_i = a_i^j \times c_j$.

%% file: oos_results_nmf.tex
\section{Evaluation}\label{sec:emp_study}
In this section, we present the evaluation procedure. We evaluate the proposed strategies on two datasets.

The first universe, see Table \ref{table:dataset1}, consists of daily returns from 2015-08-10 until 2023-03-21 of $d=21$ assets from five asset classes including cryptocurrencies with a total of 2188 daily returns for each asset and is available on \href{http://www.quantlet.de/}{quantlet.de} at \includegraphics[scale=0.05]{quantlet.png} \href{https://github.com/QuantLet/EmbeddingPortfolio/}{EmbeddingPortfolio}. It is relatively easy to get a long history of daily prices for traditional assets, but not for cryptocurrencies. Thus, we selected the cryptocurrencies with the largest market capitalization and longest history as well as the most liquid stock indices, bonds and Forex pairs. The prices are collected from the \href{https://blockchain-research-center.de/}{Blockchain Research Center (BRC)} and Bloomberg. The crypto market is a decentralized one and operates 24/7 all around the world, Forex is also decentralized and is open 24 hours a day in different part of the world but closed during weekend (typically from 4 p.m. EST on Friday until 5 p.m. EST on Sunday). Moreover, indices cannot be traded, but they can easily be replicated by future contracts or ETFs. In the empirical study, we assume that a trader replicates the various indices performances via futures, for example. The Future market is regulated and centralized with specific opening and closing time, as for the bond market. For simplicity, the price at 4 p.m. EST, which corresponds to NYSE closing time, is selected for all assets to compute the daily returns. Doing so, price variations of cryptos or Forex during NYSE closing hours are disregarded, which could correspond to the behavior of a traditional portfolio manager in the U.S. 

\begin{table}[h!]
	\small
	\centering
	\begin{tabular}{ l|c|c} 
		\hline
		\hline
		Asset class & Name & Code \\
		\hline
		\multirow{7}{4em}{Index} & FTSE EPRA & EPRA\_X \\ 
		& MSCI ACWI & MXWD\_X \\ 
		& Nikkei 225 & NKY\_X \\ 
		& Shanghai Stock Exchange Composite & SHCOMP\_X \\ 
		& SP500 & SPX\_X \\ 
		& Euro STOXX50 & SX5E\_X \\ 
		& FTSE 100 & UKX\_X \\ 
		\hline
		\multirow{4}{4em}{Bond} & U.S. 10 Year & US\_B \\ 
		& Japan 10 Year & JP\_B \\
		& UK 10 Year & UK\_B \\
		& Germany 10 Year & GE\_B \\
		\hline
		\multirow{4}{4em}{Forex} & USD/JPY & JPY\_FX \\ 
		& USD/CNY & CNY\_FX \\
		& EUR/USD & EUR\_FX \\
		& GBP/USD & GBP\_FX \\
		\hline
		Commodity & XAU/USD & GOLD\_C \\
		\hline
		\multirow{5}{4em}{Crypto} & BTC/USD & BTC \\ 
		& ETH/USD & ETH \\ 
		& DASH/USD & DASH \\ 
		& LTC/USD & LTC \\ 
		& XRP/USD & XRP \\ 
		\hline
		\hline
	\end{tabular}
	\caption{Asset universe 1 from August 2015 until March 2023}
	\label{table:dataset1}
\end{table}

Since dataset 1 has a very short history and the test set consists mostly of post COVID crash which can be considered as crisis market regime, we propose to investigate how the proposed method performs on a longer history corresponding to multiple market regimes. Alas it is impossible to test on the cryptocurrency market given its recent creation. That is why we implemented another portfolio on a second universe which consists of daily returns from 1989-02-01 until 2021-11-10 of 17 assets from three asset classes, see Table \ref{table:dataset2}. This dataset was provided by \citet{Raffinot:2017}, the missing recent prices are collected from Bloomberg data source. Finally, this dataset has 8551 daily returns over more than 30 years of history, which allows us to backtest the proposed strategy over multiple market regimes on a relatively long history. This dataset allow us to provide a fair comparison of our portfolio allocation and \citeauthor{Raffinot:2017}'s.
\begin{table}[h!]
	\small
	\centering
	\begin{tabular}{ l|c|c} 
		\hline
		\hline
		Asset class & Name & Code \\
		\hline
		\multirow{7}{4em}{Index} & FTSE 100 & FTSE \\ 
		& FTSE Small Cap & FTSE\_Small \\ 
		& MSCI ACWI & MSCI\_EM \\ 
		& SP500 & SPX\_X \\ 
		& Russell 2000 (US small cap)& Russell2000 \\ 
		& Euro STOXX50 & EuroStoxx50 \\ 
		& Euro STOXX Small Cap & EuroStoxx\_Small \\ 
		\hline
		\multirow{8}{4em}{Bond} & US 2-Year bonds& US-2Y \\ 
		& US 5-Year bonds& US-5Y \\
		& US 10-Year bonds& US-10Y \\
		& US 30-Year bonds& US-30Y \\
		& France 2-Year bonds & French-2Y \\
		& France 5-Year bonds & French-5Y \\
		& France 10-Year bonds & French-10Y \\
		& France 30-Year bonds & French-30Y \\
		\hline
		\multirow{2}{4em}{Commodity} & Gold & Gold \\ 
		& CRB Commodity Index & CRB \\ 
		\hline
		\hline
	\end{tabular}
	\caption{Asset universe 2 from February 1989 until November 2021}
	\label{table:dataset2}
\end{table}

The proposed strategies are compared with various benchmarks, such as the equally weighted allocation ($a_i = 1/d$), the equally weighted allocation per asset class ($a_i = 1/d_c$, where $d_c$ is the cardinal of the asset class as defined on Tables \ref{table:dataset1} and \ref{table:dataset2}, Markowitz portfolio \cite{Markowitz:1952} (implemented in \href{https://github.com/QuantLet/py-RobustM}{\includegraphics[scale=0.05]{quantlet.png} py-RobustM}), Robust Markowitz portfolio \cite{HKPZ:2021}  (implemented in \href{https://github.com/QuantLet/py-RobustM}{\includegraphics[scale=0.05]{quantlet.png} py-RobustM}), HRP (implemented in \cite{Prado:2016}), HCAA (implemented in \href{https://github.com/dcajasn/Riskfolio-Lib}{Riskfolio-Lib}), ERC (implemented in  \href{https://github.com/dcajasn/Riskfolio-Lib}{Riskfolio-Lib}), the S\&P 500 and the Euro Stoxx 50 indices.

\subsection{Backtesting procedure}

We compare the portfolio allocation methods in the context of target volatility strategies (TVS). Indeed, TVS have been knowing a growing interest in the last 20 years since academics and practitioners showed that they reduce downside risk by normalizing the negatively skewed return distributions \cite{Hocquard28} and tend to outperform portfolios with unconditional and static volatility in terms of risk-adjusted returns \cite{Fleming2001}. More recently, \citet{Harvey14} claim that TVS could help realize higher Sharpe ratios only for risk assets that benefit from the so-called leverage effect, such as equity or credits, and not for commodities, bonds, or currencies. However, the authors show that across all asset classes, volatility targeting reduces the likelihood of extreme returns which could lower the tail risk of multi assets portfolios. In this context, comparing the different portfolio allocations method is straightforward as it allows for a fair comparison of the strategy returns at the same unit of risk measured in terms of volatility.

In detail, we backtest the various strategies (including the single index ones) with an annualized volatility target of 5\%, a monthly rebalancing period including 2 bps transaction costs, and the estimation of the strategy parameters is calculated using the last 250 observations. The choice of this window is not discussed here and is out of the scope of this paper. Nevertheless, a yearly window is widely used in the literature on portfolio management with daily data. For this paper, we use the same dynamic target volatility strategy proposed by \citet{Jaeger:2021}. The estimation of realized volatility at time $t$ is the maximum of the portfolio volatilities measured over the previous 20 and 60 trading days, $\sigma_{t-1, 20}$ and $\sigma_{t-1, 60}$ respectively: $\widehat{\sigma}_{t-1} =  \max \left(\sigma_{t-1, 20}, \sigma_{t-1, 60}\right)$. This increases the probability that the strategy will not show higher out-of-sample volatility than the ex-ante volatility target. The leverage at the rebalancing date $t$ is then calculated as $l_t=\sigma_{\text {target }} / \widehat{\sigma}_{t-1}$, the unnormalized portfolio weights are given by:
\begin{equation}\label{eq:levered_weights}
	\widetilde{a}_{t} =  l_t \times a_{t}
\end{equation}
and the levered portfolio returns is given by:
\begin{equation}\label{eq:levered_returns}
	\widetilde{r}_{t} =  l_tr_r - (l_t -1)r_t^f
\end{equation}
where $r_t^f$ is the risk-free rate, e.g. the 4 weeks US treasury bill, thus $(l_t -1)r_t^f$ represents the cost of the target volatility strategy at time $t$.

The volatility target framework allows us to compare the various strategies for a fixed risk target, measured in terms of volatility, thus we can focus on comparing them with respect to performance measures only, described in Section \ref{sec:perf_measures}.

\subsection{Portfolio allocation summary}\label{sec:procedure}
To summarize, the main steps of the portfolio allocation procedure are:
\begin{itemize}
	\item Split the sample into a training set of length 250 days (one year) for the estimation of the parameters and a one-month test set following the training set for the out-of-sample evaluation
	\item For NMF based portfolios: select $p^*$ and estimate $W$ and $H$
	\item Compute $a_t^k$ for the portfolio $k$
	\item Compute $\widetilde{a}_t^k$ for the portfolio $k$ using Equation \eqref{eq:levered_weights}
	\item Compute the out-of-sample returns for $\widetilde{r}^k$ for the portfolio $k$ on the test set using Equation \eqref{eq:levered_returns}
\end{itemize}
Finally, repeat the operations for the next months in the test set.

\subsection{Performance evaluation}

\subsubsection{Backtest comparison}\label{sec:perf_measures}

To compare the different strategies, we use the following performance measures: 

To assess the risk-return profile of the strategies, we present the traditional Value-at-Risk ($\operatorname{VaR}$), Expected Shortfall ($\operatorname{ES}$) at 5\% level, Sharpe Ratio ($\operatorname{SR} = (\mu - r_f) / \sigma $, where $\mu$ and $\sigma$ are the sample average return and volatility),  Max drawdown ($\operatorname{MDD}$) and Calmar Ratio ($\operatorname{CR} = \mu / \operatorname{MDD}$). On top, in order to control for the skewness and kurtosis of the returns, we compute the Probabilistic Sharpe Ratio \cite{BaileyPrado:2012} that evaluates the probability that the estimated Sharpe ratio is larger than 0 in presence of non-normal returns:
$$\operatorname{PSR}=Z\left(\frac{\operatorname{SR} \sqrt{n-1}}{\sqrt{1-\gamma_{3}\operatorname{SR}+\frac{\gamma_{4}-1}{4} \operatorname{SR}^2 }}\right),$$
where  $Z$ is the cdf of the standard normal distribution, $\gamma_3$ and $\gamma_4$ are the sample skewness and kurtosis and the Minimum Track Record Length \cite{BaileyPrado:2012}:
$$\operatorname{minTRL}=1+(1-\gamma_3\operatorname{SR}+\frac{\gamma_4-1}{4} \operatorname{SR}^2)\left(\frac{Z_{\alpha}}{\operatorname{SR}}\right)^{2},$$
which indicates how long a backtest history should be to have statistical confidence that its Sharpe ratio is above 0 for a given level $\alpha$. To assess the profitability of each strategy, we compute the certainty-equivalent return defined by \citet{DeMigueletal:2009} as $\operatorname{CEQ} = (\mu - r_f) - \frac{\gamma}{2}\sigma^2$ corresponding to the risk-free rate of return that the investor is willing to accept instead of undertaking the risky strategy. $\gamma$ is the coefficient of risk aversion and results are reported for $\gamma = 1$. The CEQ represent the level of expected utility of a mean-variance investor. We also compute the total average turnover per rebalancing: $$\operatorname{TTO} = 1/T\sum_{t=1}^{T}\lVert a_t - a_{t^-}\rVert_1$$ where $t$ is the rebalancing date and $t^-$ is the last date before rebalancing. Finally, to measure the difficulty of implementing the volatility target strategy, we present the average leverage factor $l_t$ for each strategy which shows how far the original strategies are from the 5\% annual volatility target

To quantify diversification in the portfolios, we introduce the following metrics:

We propose first to use a simple metric, the average Sum of Squared Portfolio Weights per rebalancing period (SSPW) \cite{GoetzmannKumar:2008} that shows the underlying level of diversification in a portfolio at the asset level assuming no factor model for the returns. The simple SSPW is defined as $\mathcal{H}_t(a) = \sum_{i=1}^{d} a_{i, t}^{2}$. This is also called the Herfindahl index applied to the portfolio weights. It takes the value 1 for a portfolio with maximum concentration in one asset ($\exists!\ i, a_i = 1$ and $\forall j \neq i, a_j = 0$) and $1/d$ for a portfolio with uniform allocation. We consider the normalized index to scale the statistics onto $[0,1]$: $\mathcal{H}_t^*(a)= \left(d\ \mathcal{H}_t(a) - 1\right)/(d - 1)$ and average it over the rebalancing periods to obtain:
\begin{equation}\label{eq:sspw}
	\operatorname{SSPW}(a) = 1/T \sum_{t=1}^{T} \mathcal{H}_t^*(a)
\end{equation}

For factor models, the classical method to measure diversification is to decompose the risk using the Euler principle and compute the risk contribution of each risk factor to the total portfolio risk. We introduce the average number of NMF bets (denoted NMF Bets), a variant of the effect number of bets from \citet{meucci2009managing}, that shows the degree of diversification in terms of the obtained NMF factors and their marginal risk contributions to the overall portfolio volatility. Theorem 2 from \citet{roncalli2016risk} explains how to compute it for a portfolio allocation $a$ with volatility $\sigma(a) = \sqrt{a^\top \Sigma a}$. Let us denote $\mathcal{RC}(z_k)$, the risk contribution of the NMF factors, and $\mathcal{RC}(y_k) $, the risk contribution of the additional factors included in the model residuals $\eta$ in Equation \ref{eq:nmf_model}, for a portfolio allocation $a$ we have:
\begin{align}
	\mathcal{RC}(z_k) &= \frac{\left(\widetilde{W}a\right)_j    \left(\widetilde{W}^+ \Sigma a\right)_j }{\sigma(a)},\quad 1\leq k \leq p \label{eq:rc_z}\\
	\mathcal{RC}(y_k) &= \frac{\left(\widetilde{V}a\right)_j    \left(\widetilde{V}^+ \Sigma a\right)_j }{\sigma(a)},\quad p + 1 \leq k \leq d \label{eq:rc_y}
\end{align}
where $\widetilde{V}$ is a matrix that spans the left null space of $V = \widetilde{W}^\top$. We can verify that $\sigma(a)  = \sum_{i=1}^p \mathcal{RC}(z_k) + \sum_{i=1}^p \mathcal{RC}(y_k)$. $\mathcal{RC}(z_k)$ represents the risk contributions of each NMF factor, and $\mathcal{RC}(y_k)$ represents the risk contributions of the idiosyncratic risk of the portfolio allocation $a$. If the returns are fully explained by the $p$ NMF factors, then a portfolio allocation that is well balanced over the factors, must have $\operatorname{NMF\ Bets} \approx p$. In general, for the NMF-based portfolios $\operatorname{NMFRP}$ and $\operatorname{NMFRB}$, we should have $\operatorname{NMF\ Bets} \geq p$ for each rebalancing period since the idiosyncratic risk associated with the additional factors $y_k$ is not 0. However, if we observe $\operatorname{NMF\ Bets} \gg p$, the NMF factors probably do not represent the portfolio well, and the risk is spread in some additional factors. Finally, we introduce the probability distribution:
\begin{equation}\label{eq:rc_nmf}
	m_i(a) = \lvert \mathcal{RC}_i \rvert / \sum_{k=1}^d \lvert \mathcal{RC}_k \rvert,\quad \forall 1\leq i \leq d
\end{equation} and we define the number of NMF bets for a portfolio by taking the Shannon entropy:
\begin{equation}
	\operatorname{NMF\ Bets}(a) =  \exp \left(- \sum_{i=1}^d m_i(a) \log m_i(a)\right)
\end{equation}

In factor models, it is natural to decompose the risk using the risk contribution of each NMF factor to the total portfolio risk. However, such contributions are spurious, because, in reality, they contain effects from all the factors at once, they can be negative which requires taking the absolute value and the NMF factors might be correlated. \citet{Meucci2015RiskBA} proposed to find the set of uncorrelated factors that are optimized to track the original factors used to allocate the portfolio, named the "Minimum-Torsion Bets". The authors introduce a novel diversification metric called the "Effective Number of Minimum-Torsion Bets" ($\operatorname{MT\ Bets}$) which summarizes in one number the structure of diversification contained in the set of uncorrelated bets in the portfolio, instead of using the marginal contributions of the original correlated factors. Before defining $\operatorname{MT\ Bets}$, we must explicitly express the full set of $d$ factors $F$ given by the NMF model from Equation \eqref{eq:nmf_model} and the portfolio weights in the full NMF factors $h_t$. We refer the reader to the previously mentioned papers, we have:
\begin{align}
	F_t & = \begin{pmatrix}
		z_t\\
		y_t
	\end{pmatrix} = \begin{pmatrix}
		z_t\\
		\left(\widetilde{V}^+\right)^\top r_t 
	\end{pmatrix} \\
	\widetilde{h}_t &= \begin{pmatrix}
		\widetilde{a}_t\\
		\widetilde{a}_t^f
	\end{pmatrix} = \begin{pmatrix}
		\widetilde{W}^\top a\\
		\widetilde{V} a
	\end{pmatrix}
\end{align}
where  $z_t$ is the $p$ vector of latent NMF factors and $y_t$ is the $d-p$ vector of additional factors from the model residuals $\eta_t$. We assume $Cov(z_i, y_j) = 0, \forall 1\leq i \leq p,\quad  p+1\leq j \leq d$.

Let us denote $\mathbf{t}_F$ the minimum-torsion rotation for the factors $F$, we have $\mathbf{t}_F = dg(\Omega_F)^{1/2}\pi c^{-1}dg(\Omega_F)^{1/2}$, where $dg(\Omega_F)^{1/2}$ is the diagonal matrix of standard deviation of the factors $F$. The final risk contributions of a portfolio allocation $a$ can be expressed in terms of the Minimum-Torsion Bets to the factors:
$$p_i(a) = \frac{\bigl({\mathbf{t}_F^\top}^{-1}\widetilde{h}\bigr) \circ \bigl(\mathbf{t}_F\Omega_F \widetilde{h}\bigr)}{\widetilde{h}^\top\Omega_F\widetilde{h}}$$
where $\circ$ denotes the term-by-term product.
The Minimum-Torsion Bets contributions represent uncorrelated bets and define a probability distribution, thus they can effectively be interpreted as contributions from truly separate sources of risk.
The Effective Number of Minimum-Torsion Bets is then given by taking the negative Shannon entropy over the $p_i$ we have:
\begin{equation}\label{eq:mt_bets}
	\operatorname{MT\ Bets}_F(a) = \exp \left(- \sum_{i=1}^d p_i(a) \log p_i(a)\right)
\end{equation}
$\operatorname{MT\ Bets}_F$ represents the true uncorrelated bets of a portfolio $a$ with respect to the NMF factors. If the NMF factors have a quasi diagonal-correlation matrix and represent well the asset returns ($\forall i,\ p+1\leq i \leq d,\ p_i(a) \approx 0$), then a well diversified portfolio must have $\operatorname{MT\ Bets}_F \approx p$. If $\operatorname{MT\ Bets}_F > p$, then the idiosyncratic factors have significant additional uncorrelated sources of risk which help diversify the portfolio. Finally, if $\operatorname{MT\ Bets}_F < p$, then probably the NMF factors $z$ are correlated or some idiosyncratic factors $y$ are correlated with the NMF factors $z$ and the NMF factors are not well diversified.

We also compute the $\operatorname{MT\ Bets}(a)$ in the original assets space. It allows to compare the degree of diversification of a portfolio $a$ between the factors, which represent synthetic assets classes and within, at the asset level. Replacing $F$ with $R$ the original assets returns and $\mathbf{t}_F$ with $\mathbf{t}$ the minimum-torsion rotation for $R$, we can express the $\operatorname{MT\ Bets}(a)$ in the original asset space with:
$$p_i(a) = \frac{\bigl({\mathbf{t}^\top}^{-1}a\bigr) \circ \bigl(\mathbf{t} \Sigma a\bigr)}{a^\top\Sigma a}$$
$\operatorname{MT\ Bets}$ measures how well a portfolio is diversified concerning the original assets. It is interesting to compare $\operatorname{MT\ Bets}_F$ with $\operatorname{MT\ Bets}$ for a given portfolio $a$. Indeed, since the factors dimension represents $p < d$ synthetic asset classes, then $\operatorname{MT\ Bets} > \operatorname{MT\ Bets}_F$ means that the portfolio is well diversified between the synthetic asset classes and within. If we observe the opposite, then the portfolio is more diversified between the assets classes than within. A very well diversified portfolio should be diversified at both level, having $p \leq \operatorname{MT\ Bets}_F \ll \operatorname{MT\ Bets} \leq d $.

\subsubsection{Statistical evaluation}

To evaluate the predictive performance of the NMF model for the assets stock returns, we calculate on the test set, the total $R^2$ defined by \cite{Kelly:2018} and \cite{gu2020empirical} as:
\begin{equation}\label{eq:total_r2}
	R_{total}^2 = \frac{ \sum_{i,t}(r_{it} - \widehat{r}_{it} )}{ \sum_{i,t} r_{it}^2}
\end{equation}
where $t$ does not belong to the train set. This metric uses as a denominator the sum of squared returns without demeaning because the historical mean of stock returns is too noisy and artificially improves the performance evaluation. The benchmark used for the $R_{total}^2$ is then a prediction value of zero. The $R_{total}^2$ quantifies the explanatory power of the factor realizations, and thus assesses the model's description of individual stock riskiness.

\section{Empirical Results}\label{sec:empirical_results}
\subsection{Backtest}

\paragraph{Global cryptocurrency portfolio}

\definecolor{equal}{HTML}{808080}
\definecolor{equalclass}{HTML}{7F3417}
\definecolor{markowitz}{HTML}{F6C2CB}
\definecolor{robustm}{HTML}{2C5658}
\definecolor{HRP}{HTML}{6D157C}
\definecolor{NMFRP}{HTML}{3601FF}
\definecolor{HCAA}{HTML}{F602FF}
\definecolor{NMFRB}{HTML}{15E943}
\definecolor{ERC}{HTML}{30F1FF}
\definecolor{SP500}{HTML}{DF0103}

\definecolor{better}{HTML}{BA141A}
\definecolor{worse}{HTML}{F6AA8B}

\begin{figure}[!ht]
	\includegraphics[width=\textwidth]{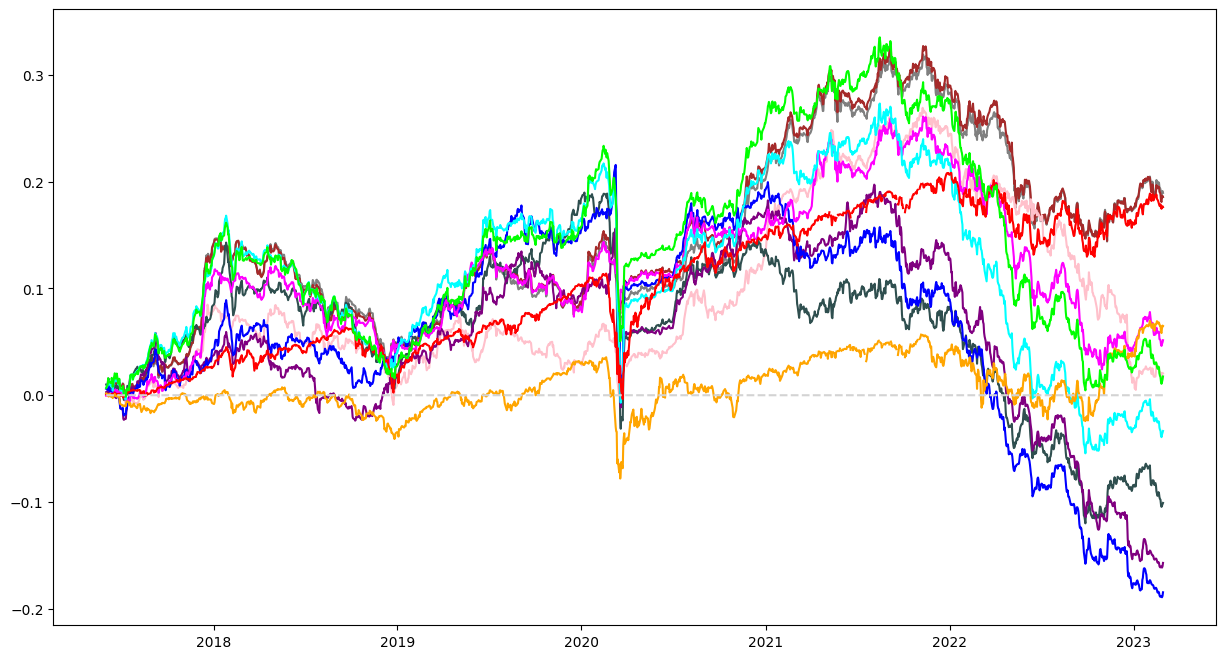}
	\centering	
	\caption{Total return over the test set of   \textcolor{NMFRP}{NMFRP},  \textcolor{NMFRB}{NMFRB}, \textcolor{HRP}{HRP},  \textcolor{ERC}{ERC},  \textcolor{HCAA}{HCAA}, \textcolor{markowitz}{Markowitz}, \textcolor{robustm}{robust Markowitz}, \textcolor{equal}{equal}, \textcolor{equalclass}{equal class}, \textcolor{SP500}{SP500}, \textcolor{Orange}{EuroStoxx50} weighted portfolios \quantletNMFRB} 
	\label{fig:port_perf_all_dataset1}
\end{figure}

Results of the historical backtest on dataset 1 are summarized in the Table \ref{table:stats_heatmap_dataset1} and cumulative performances represented in Figure \ref{fig:port_perf_all_dataset1}. It is very hard to beat naive portfolios and benchmark indices in terms of total and risk-adjusted returns on this investment universe, possibly because the diversification over the asset universe during the COVID crisis in March 2020 and the momentum effects from the cryptocurrency market after the crash are quite strong. On top, all strategies, except Equal and Equal class, have a $\operatorname{minTRL}$ too large compared to the size of the datasets, indicating that the historical sample is not long enough to be confident at 95\% of obtaining a positive average return with these strategies.

Inverse-variance-based allocation methods, except Markowitz, e.g. HRP, NMFRP, and Robust-M, show poor performance concerning risk-adjusted return, with an average $\operatorname{PSR}$ of 0.15, 0.10, and 0.28 respectively only, and a large $\operatorname{minTRL}$. This is explained by the large leverage factor required to achieve the volatility target which comes with a greater cost (see Equation \eqref{eq:levered_returns}). On top, HRP and NMFRP do not achieve diversification at the asset level, having the largest $\operatorname{SSPW}$ (0.53 and 0.25 on average, respectively) after Markowitz's portfolio (0.75). This is also reflected in the small $\operatorname{MT\ Bets}$ for HRP (5.32 out of 21 assets on average). This causes a large concentration of capital in low volatility assets, such as bonds, which did not perform well during most the period of study, as shown in the weights bar chart on Figure \ref{fig:weightshrp_1} in Appendix, and increases the cost of leverage. On the other hand, inverse variance-based allocation methods achieve good diversification at the NMF factors level. HRP makes on average 8.88 NMF bets (the largest number observed) and 8.78 $\operatorname{MT\  bets}_F$. This means that HRP is well diversified between synthetics asset classes but not between the original assets. Indeed, the main drawback of these two methods is that they do not diversify the risk well when applied to such investment universes where assets show different volatility structures.
\begin{table}[!ht]
	\includegraphics[width=\textwidth]{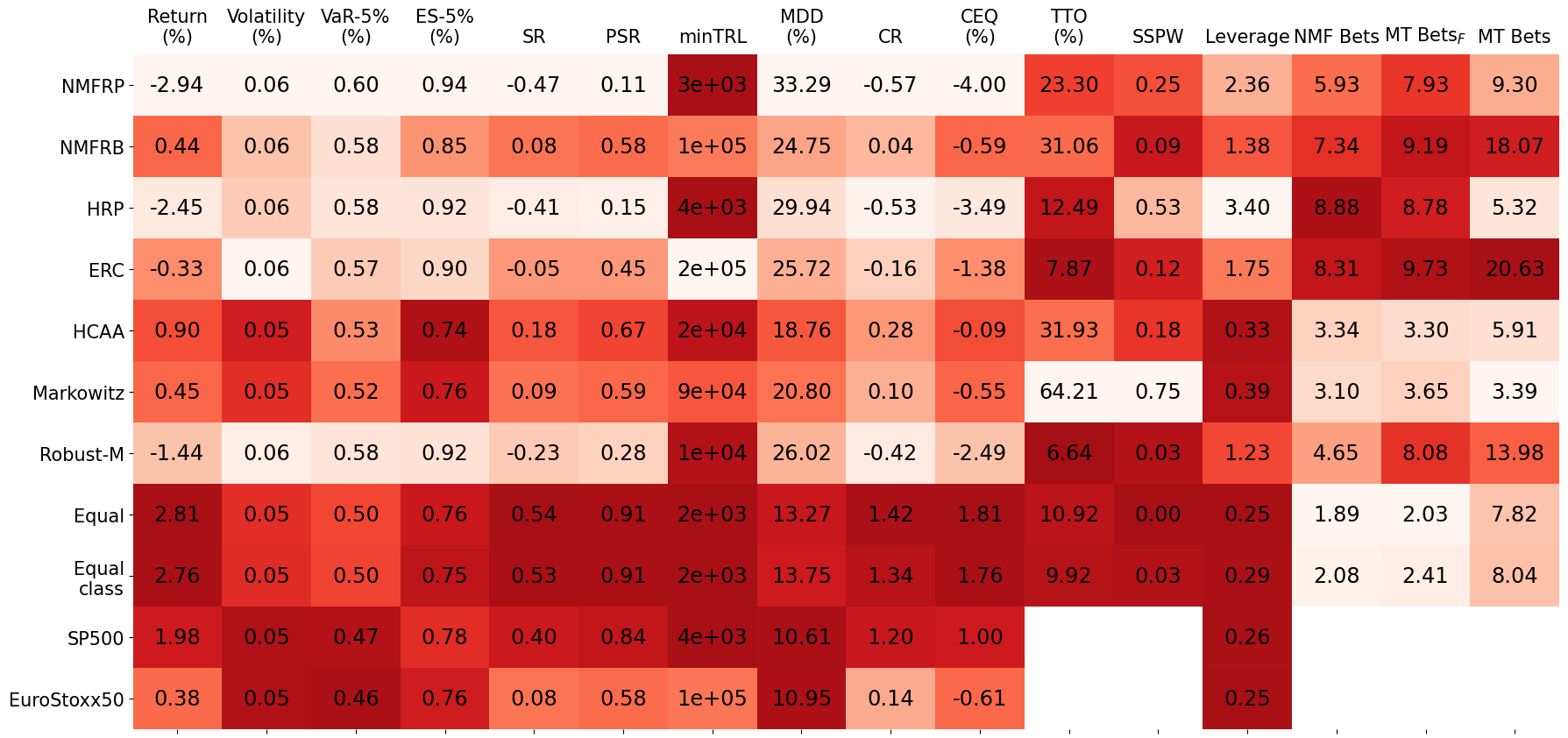}
	\centering
	\caption{Backtest statistics comparison on dataset 1. A better performance is indicated by a darker red color: \textcolor{better}{SR=2 is better} than \textcolor{worse}{SR=1} \quantletNMFRB}
	\label{table:stats_heatmap_dataset1}
\end{table}

On this dataset, the proposed method $\operatorname{NMFRB}$ seems to outperform inverse-variance-based allocations in terms of total and risk-adjusted return with a larger average return, SR, and PSR (0.44, 0.08, and 0.58 respectively). We investigate this further with a test of superior predictive ability in the next section \ref{sec:data_snooping}. As the ERC allocation, it achieves good diversification both at the asset level with $\operatorname{SSPW} = 0.09$, which is lower than the $\operatorname{SSPW}$ for HRP or HCAA and most importantly, 18.07 $\operatorname{MT\ bets}$ which is close to the upper bound 21. It is also well diversified at the factor level with 7.34 $\operatorname{NMF\ bets}$, almost double the average number of factors (see Figure \ref{fig:n_factors1}) and 9.19 uncorrelated $\operatorname{MT\ bets}_F$. Thanks to this larger inherent diversification, the required leverage, 1.38 on average, is relatively lower than for HRP, NMFRP or Robust-M and the volatility target is cheaper to achieve. In general, NMFRB seems to be well diversified between synthetic assets classes and within. On top, the NMFRB strategy has a relatively lower tail risk compared to HRP with a smaller Expected Shortfall (0.85 and 0.92, respectively) and MDD (24.75 and 29.94, respectively). 
 
Finally, it is worth noticing that the HCAA approach does not achieve a good diversification at both asset and factor levels with, on average, relatively low NMF, $\operatorname{MT}_F$, and $\operatorname{MT\ bets}$ (3.34, 3.30 and 5.91, respectively). Indeed, the weights in Figure \ref{fig:weightshcaa_1} show that HCAA as a relatively large concentration in the cryptocurrencies and a few large outliers compared to other portfolios. As expected, Markowitz's portfolio presents the worst SSPW (0.75) and largest turnover (64.21\%) indicating a large and unstable concentration in a few assets as presented in Figure \ref{fig:weightsmarkowitz_1}; on the other hand, the naive Equal and Equal Class allocations have a good diversification at the asset level by definition and makes more uncorrelated bets than HRP, HCAA, and Markowitz portfolios, but the worst diversification between the synthetic asset classes with only 2.03 and 2.41 $\operatorname{MT\ bets}_F$ respectively. Finally, those strategies put more weights on the assets with larger volatility, thus they have a larger inherent volatility which considerably reduces the required leverage to only a fraction of the invested capital (33, 39, 25 and 29\% for HCAA, Markowitz, Equal and Equal class strategies respectively). Thus, they benefit from investing in the risk free rate.

In summary, in this universe, the proposed NMFRB method outperforms naive and hierarchical-clustering-based strategies in terms of diversification. However, it does not outperform naive strategies in terms of risk-adjusted returns.
 
\begin{figure}[!ht]
	\includegraphics[width=\textwidth]{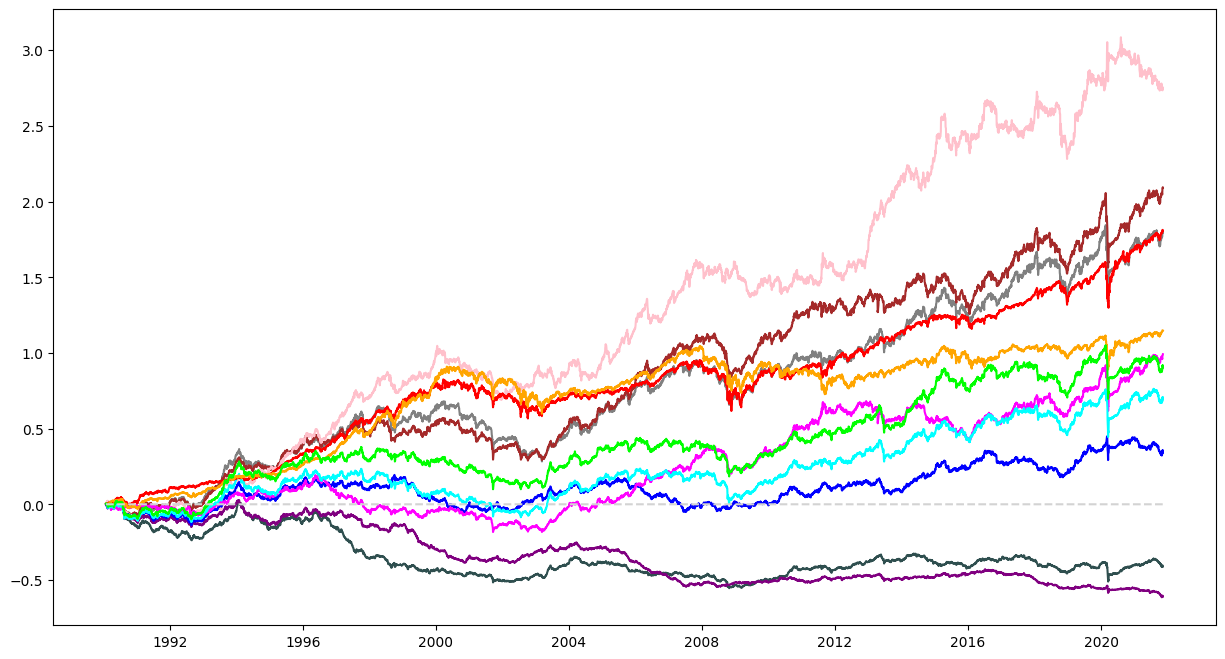}
	\centering	
	\caption{Total return over the test set of   \textcolor{NMFRP}{NMFRP},  \textcolor{NMFRB}{NMFRB}, \textcolor{HRP}{HRP},  \textcolor{ERC}{ERC},  \textcolor{HCAA}{HCAA}, \textcolor{markowitz}{Markowitz}, \textcolor{robustm}{robust Markowitz}, \textcolor{equal}{equal}, \textcolor{equalclass}{equal class}, \textcolor{SP500}{SP500}, \textcolor{Orange}{EuroStoxx50} weighted portfolios \quantletNMFRB}
	\label{fig:port_perf_all_dataset2}
\end{figure}

\paragraph{Global traditional portfolio}
Figure \ref{fig:port_perf_all_dataset2} and Table \ref{table:stats_heatmap_dataset2} show the strategies performance on dataset 2. 
First, we notice that the unlevered Equal, Equal class, and HCAA portfolios are fairly aligned with a the 5\% annual volatility target, as their average leverage factor is near 1 on average. Markowitz portfolio has the largest variance which implies lower leverage, thus the corresponding levered strategy profits from risk-free returns. HRP portfolio has the smallest volatility unlevered, thus it requires on average a 5.63 leverage factor which incures significant borrowing costs. On the other hand, the leverage of the NMFRB strategy is located in-between, at 2.55 on average.

The historical sample seems long enough for all strategies except for the NMFRP, Robust-M, and EuroStoxx50 to be confident at 95\% of obtaining a positive average return with these strategies. The proposed NMFRB strategy performs at least as well as all other strategies in terms of risk-adjusted returns, having a similar average return, tail risk profile, and SR and PSR, but a much lower MDD (20.26\%) implying a much larger CR (4.43), except for the naïve and Markowitz portfolios. It outperforms the HRP strategy in terms of risk-adjusted returns (see Section \ref{sec:data_snooping}). On average, NMFRB is well diversified at both the asset and factor levels with SSPW=0.10 and making 14.59 MT Bets (which is close to the upper bound 17 and larger than the other portfolios, except for ERC), 4.97 and 5.57 NMF and $\operatorname{MT}_F$ bets, respectively, which is larger than the other portfolios, except for ERC and Robust-M. In general, NMFRB seems to be well diversified between synthetic asset classes and within.
\begin{table}[!ht]
	\includegraphics[width=\textwidth]{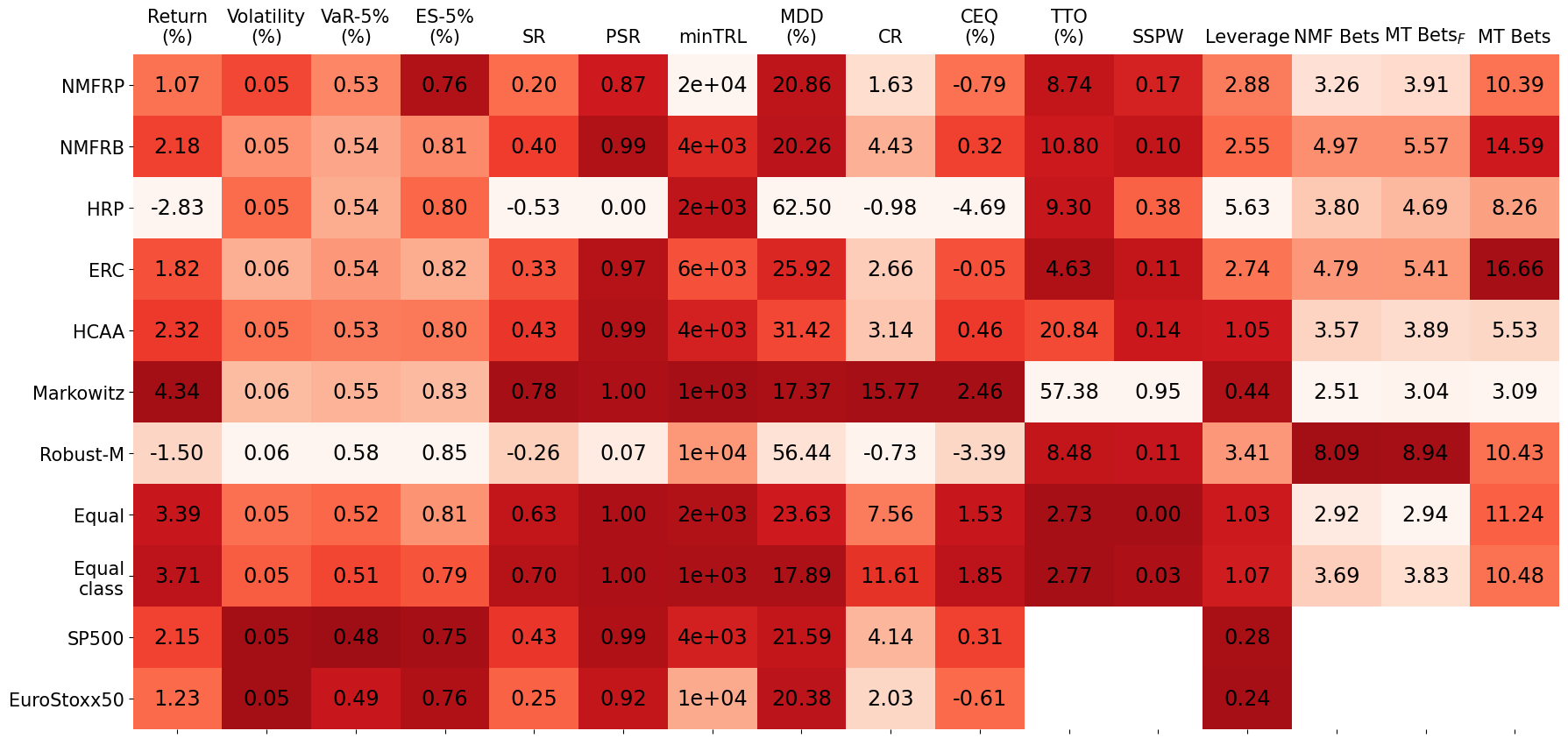}
	\centering
	\caption{Backtest statistics comparison on dataset 2. A better performance is indicated by a darker red color: \textcolor{better}{SR=2 is better} than \textcolor{worse}{SR=1} \quantletNMFRB}
	\label{table:stats_heatmap_dataset2}
\end{table}

It is worth noticing that on both datasets, while Markowitz allocation achieves a good risk-adjusted return performance, it does not provide any diversification effect. Instead, it luckily picked an asset with a good average return. On the other hand, while the Robust Markowitz method produces a well-diversified portfolio, it underperforms most of the other strategies in terms of risk-adjusted return. Similarly, on both datasets, the HCAA approach does not achieve good diversification at both asset and factor levels and has a large turnover.

In short, the NMFRB portfolio outperforms HRP portfolio on dataset2 in terms of expected and risk-adjusted returns controlled for non-normality and for diversification. This indicates that, for a Markowitz-type investor, the proposed strategy should be preferred even if it allocates more weight to assets with larger volatility compared to HRP (see Figure \ref{fig:weights_2}). This may be the outcome of diversifying the allocation between and within synthetic asset classes that exhibit little or no correlation. We cannot make such a strong statement for the investment universe 1, despite indications suggested by sample estimates of various performance statistics, due to insufficient historical data as we explain in the next subsection.

\subsection{Data snooping}\label{sec:data_snooping}

Data snooping can be defined as the act of analyzing data repeatedly until a significant result is found, without proper accounting for the multiple comparisons made \cite{white2000reality}. Also called P-Hacking or HARKing in other scientific disciplines, see \href{https://www.quantinar.com/course/35/phacking}{\quantinar P-Hacking}, such a practice can lead to false positives and spurious conclusions, as the probability of finding a significant result purely by chance increases with the number of comparisons made. To avoid data snooping, we perform a test for Superior Predictive Ability (SPA) as in example 3 of \citet{10.2307/27638834}. It tests whether the expected returns of any strategy from a set of alternatives is better than the expected returns from a benchmark strategy. This procedure accounts for dependence between the strategy returns and the fact that there are potentially alternative strategies being considered.

In our context, we test whether the strategies outperform the $\operatorname{HRP}$ portfolio in terms of expected returns. We apply the test to the relative out-of-sample portfolio loss series $-r_t^{\operatorname{HRP}} + r_t^{k}$, where $k$ is any other portfolio than $\operatorname{HRP}$. We apply the test on two sets of alternatives, one containing all strategies except HRP (denoted Full set) and one containing only the NMF based strategies, NMFRP and NMFRB (denoted NMF set).

\begin{table}[h!]
	\small
	\centering
	\begin{tabular}{ c | c | c | c | c} 
		\hline
		\hline
		Universe & Set & $p_{lower}$ & $p_{consistent}$ & $p_{upper}$ \\
		\hline
		\multirow{2}{4em}{Dataset1} & Full set & \textbf{4.3} & \textbf{4.3} & \textbf{4.3} \\ 
													  & NMF set & $8.3$ & $8.4$ & $8.4$ \\ 

		\multirow{2}{4em}{Dataset2} & Full set & \textbf{0.0} & \textbf{0.0} & \textbf{0.0} \\
		& NMF set  & \textbf{0.0} & \textbf{0.0} & \textbf{0.0} \\
		\hline
		\hline
	\end{tabular}
	\caption{p-values (of order $10^{-2}$) for the SPA test over HRP portfolio on dataset 1 and 2, \textbf{bold p-values are significant at level 0.05} \quantletNMFRB}
	\label{table:spa}
\end{table}
The p-values for the SPA test on dataset 1 and 2 are represented in table \ref{table:spa}. On dataset 1, the results are unclear for the proposed strategies: we reject the null hypothesis at level 5\% for the Full set, but cannot reject it for the NFM set at the same level. We can only accept the superior predictive ability of the NMF set over HRP at the level 10\%. This weak evidence of superiority might be explained by the short sample size of the dataset1. Indeed, on dataset 2 which has a much longer history, all p-values are significant at the 0.05 level which is strong evidence that the NMF-based strategies outperform $\operatorname{HRP}$ in terms of expected returns, if the history is long enough. Since, we are backtesting the strategies with a constant volatility target and observing that the strategies have a similar volatility out-of-sample, we can say that the target-volatility NMF-based strategies outperform $\operatorname{HRP}$ in terms of risk-adjusted returns. We also performed a SPA test using the HCAA portfolio as a baseline. Alas, we cannot reject the null hypothesis for the NMFRP or NMFRB strategies in that case.

\subsection{Risk exposures analysis}

The main advantage of the NMF factors over other factorization models is that we can easily interpret them as synthetic asset classes representing specific long-only portfolios and cluster centroids. Nevertheless, since the nature of the factors varies across the rebalancing periods, it is difficult to study them over time. Thus, we summarize the information in this section with various figures and we study the factors at specific dates to give example of interpretation and illustrate their behavior.

\subsubsection{Interpretable NMF factors}

\begin{figure}[ht!]
	\centering
	\vskip 0pt
	\includegraphics[width=\linewidth]{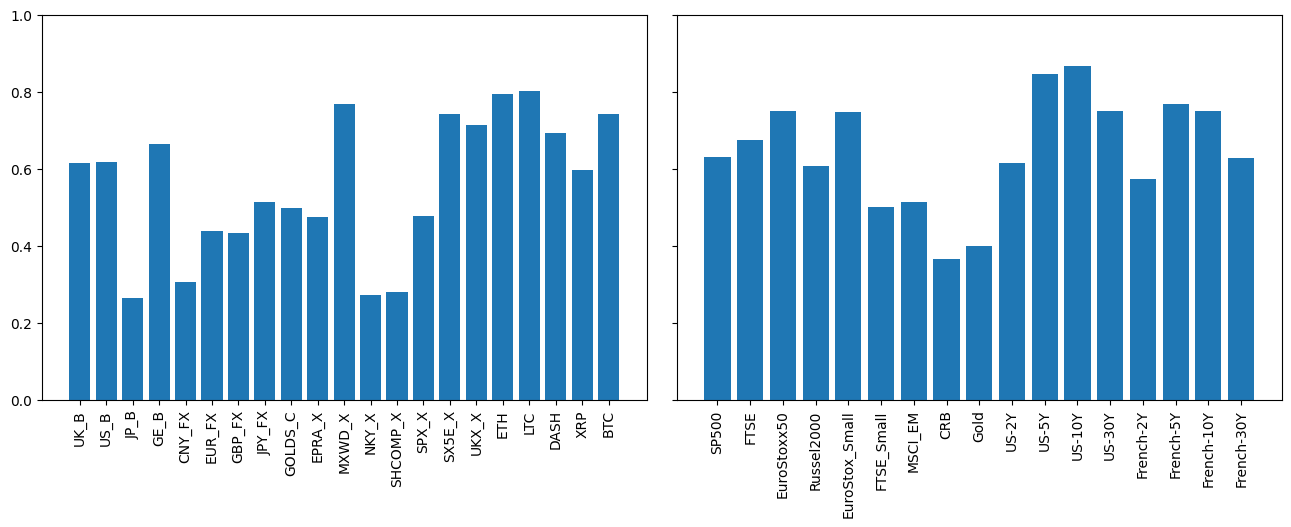}
	\caption{Monthly average $R_{total}^2$ for each assetson dataset 1 (left) and 2 (right) \quantletNMFRB}
	\label{fig:total_r2}
\end{figure}

Before interpreting the NMF model, we evaluate its predictive performance for the assets stock returns with the average of the total $R^2$ (see Equation \eqref{eq:total_r2}) over the different rebalancing periods, shown in Figure \ref{fig:total_r2}. 

In the first dataset, the average $R_{total}^2$ is 55\%. The factors explain the crypto assets well, with $R_{total}^2$ values of 80, 80, 70, and 74\% for ETH, LTC, DASH, and BTC, respectively. For the stock market, MXWD\_X, SX5E\_X, and UKX\_X indices are better explained with $R_{total}^2$ values of 77, 74, and 71\%, respectively. However, the UK, US, and GER bonds have $R_{total}^2$ values slightly above 60\%. The Asian region is, on average, poorly explained by the NMF factors, as evidenced by the Japanese bond, CNY/USD, NKY/USD pairs, and Shanghai Stock Exchange composite index, which have the lowest $R_{total}^2$ values below 30\%.


The NMF factors on the second dataset provides a better explanation of the assets compared to the first one with an average $R_{total}^2=65$\%, 10 points higher than the first dataset. The American and French bonds with mid-term tenors of 5 and 10 years are the best explained assets in the second dataset. The $R_{total}^2$ value for these assets is approximately 80\%, indicating a high level of explanatory power. However, the commodities represented by Gold and CRB index are poorly explained by the NMF factors.

\begin{figure}[ht!]
	\centering
	\includegraphics[width=\linewidth]{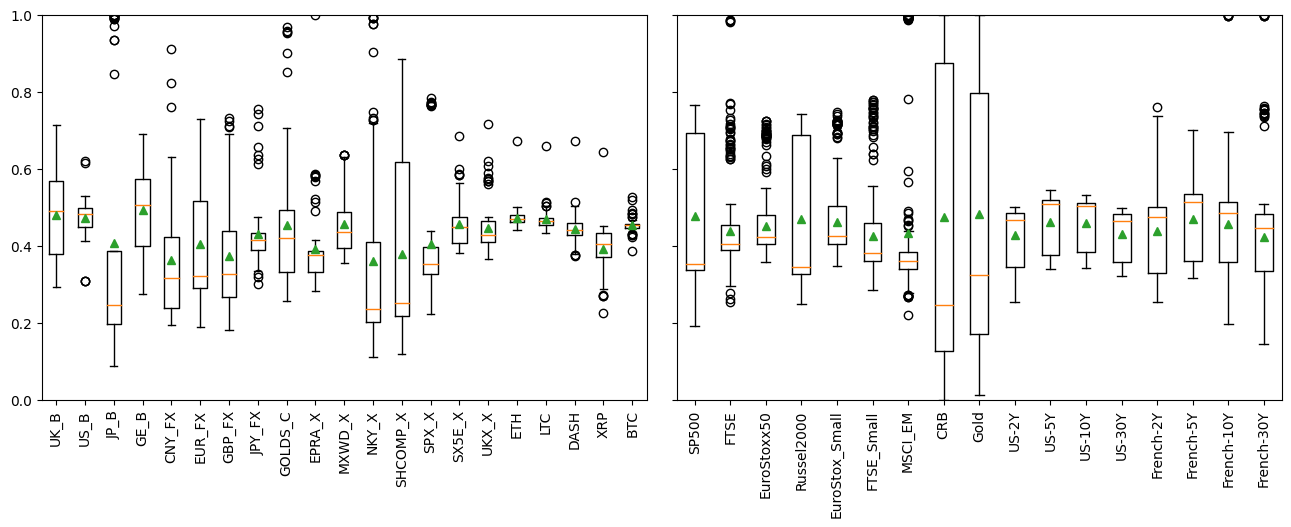}
	\caption{Max factor loading ($\max_k W_{:k}$) for each asset on dataset 1 (left) and 2 (right) \quantletNMFRB}
	\label{fig:assets_loading}
\end{figure}

\begin{figure}[ht!]
	\centering
	\begin{subfigure}[t]{0.49\textwidth}
		\centering
		\includegraphics[width=\textwidth]{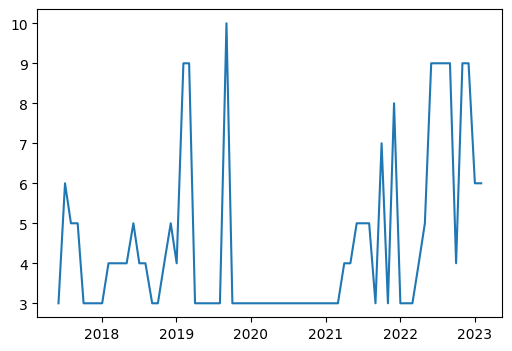}
		\caption{Dataset1}
		\label{fig:n_factors1}
	\end{subfigure}
	\hfill
	\begin{subfigure}[t]{0.49\textwidth}
		\centering
		\includegraphics[width=\textwidth]{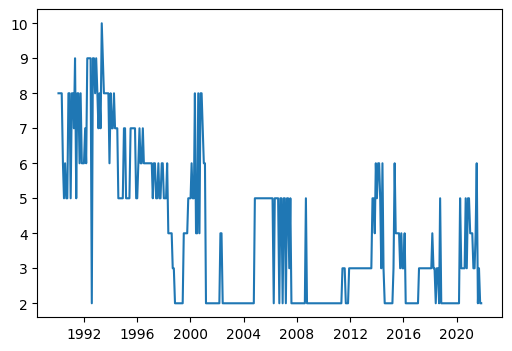}
		\caption{Dataset 2}
		\label{fig:n_factors2}
	\end{subfigure}
	\caption[]{Number of NMF factors over the rebalancing periods \quantletNMFRB}
	\label{fig:n_factors}
\end{figure}

Figure \ref{fig:assets_loading} presents the distribution of maximum factor loadings for each asset ($\forall i,\ \max_k W_{ik}$) across the rebalancing periods. This visualization helps explain the observed discrepancies in the $R_{total}^2$. Specifically, the assets that are poorly explained in the analysis have factor loadings with larger variance, indicating that the corresponding factors are not stable across the rebalancing period, and smaller median, indicating a lower correlation between the assets and the corresponding explaining factor. Conversely, well-explained assets have factor loadings with relatively smaller variances, such as ETH, LTC, and BTC in dataset 1, indicating that the corresponding factors are stable over the rebalancing period, and larger median, expressing a larger correlation between asset and factor.
In summary, the discovered factors are more suitable for the analysis of bonds, stocks and cryptocurrencies, but it may not provide adequate insights for Asian assets in the first dataset and commodities in the second dataset,

We represent the number of risk factors over time on both datasets in Figure \ref{fig:n_factors}. For dataset 1, it oscillates between a minimum of 3 and a maximum of 10 with an average of 4 and between 2 and 10 with an average of 4 as well for the second dataset. First, we notice that the model with one factor only has never been selected on both datasets. It suggests that, at least 2 and 3 clusters exist within the correlation matrix, for dataset 2 and 1 respectively, and could help diversify the portfolio. In particular, it seems that there are 3 consistent factors in dataset 1 mostly representing the bond, crypto, and stock markets respectively as we explain in the next paragraph. In dataset 2, at least 2 factors seem to explain the correlation of the assets, one representing mostly the assets with low volatility (e.g. the bond market) and the other representing assets with larger volatility, e.g. stocks or commodities. On top, we never selected a model with $p \approx d$ which also indicates that assets can be clustered and that some of them might be redundant. On both datasets, we observe that the number of risk factors switches from stable periods (e.g. 2020 and 2008 for datasets 1 and 2, respectively) to unstable ones (e.g. 2022 and 2013-2021 for datasets 1 and 2, respectively). The stability of the number of factors could increase when fitting the NMF model on a larger historical window. In periods of crisis or post-crisis the number of factors drops to its minimum, 3 factors for the universe 1 during the COVID crisis in 2020 and 2 for the second one after the "dotcom" bubble in the early 2000s and during the "subprime" crisis from the end of 2007 until 2011. This behavior confirms the stylized fact that assets tend to be more correlated during crisis which reduces the number of underlying low-correlated factors and their diversification effect. 

\begin{figure}[ht!]
	\centering
	\begin{subfigure}[b]{0.49\textwidth}
		\centering
		\includegraphics[width=\linewidth]{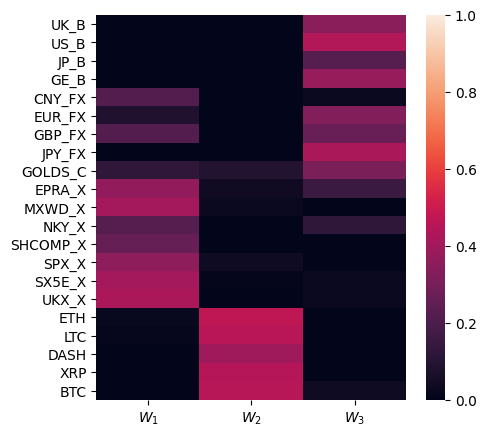}
		\caption{NMF factor loadings during the COVID crisis in May 2020 in universe 1}
		\label{fig:exloadingscovid}
	\end{subfigure}
	\hfill
	\begin{subfigure}[b]{0.49\textwidth}
		\centering
		\includegraphics[width=\linewidth]{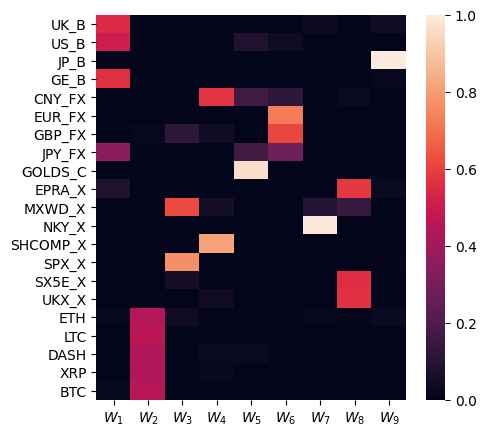}
		\caption{NMF factor before Fed's rate increase in December 2021 in universe 1}
		\label{fig:exloadingsendlowrate2022}
	\end{subfigure}
	\vskip\baselineskip
	
	\begin{subfigure}[b]{0.49\textwidth}
		\centering
		\includegraphics[width=\linewidth]{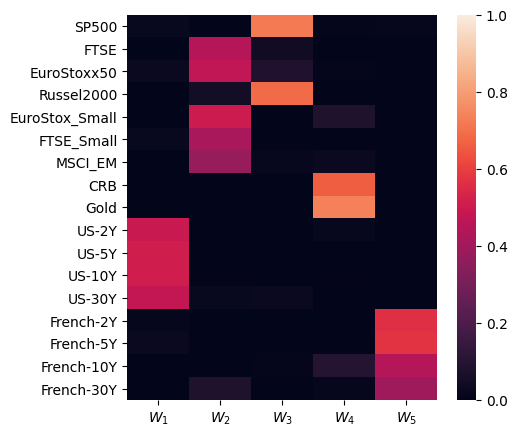}
		\caption{NMF factor loadings before the dotcom bubble crash in March 2000 in universe 2}
		\label{fig:exloadingsbeforecrisis2}
	\end{subfigure}
	\hfill
	\begin{subfigure}[b]{0.49\textwidth}
		\centering
		\includegraphics[width=\linewidth]{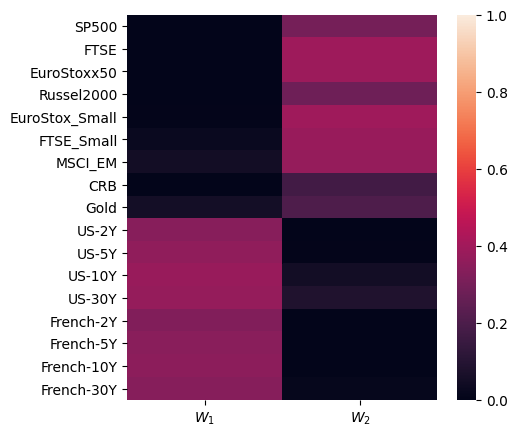}
		\caption{NMF factor loadings during 2008 crisis in January 2008 in universe 2}
		\label{fig:exloadings20082}
	\end{subfigure}
	\caption[]{NMF factor loadings example \quantletNMFRB}
	\label{fig:ex_oos_loadings}
\end{figure}

Some examples of factor loadings are given in Figure \ref{fig:ex_oos_loadings}. When NMF finds 3 factors on dataset 1, as in Figure \ref{fig:exloadingscovid} representing the factor loadings obtained during the COVID crisis, we can interpret them as 3 long-only portfolios: Factor 1, 2, and 3 are respectively exposed to the global stock markets including also CNY/USD Forex pair, the crypto market, and the bond and Forex market, including also Gold. When more factors are included, as in Figure \ref{fig:exloadingsendlowrate2022} with 8 factors, the former two global factors (factors 1 and 3) are split while the crypto factor keeps its integrity. This split can also be interpreted: for example, in Figure \ref{fig:exloadingsendlowrate2022}, we observe that the bond market is split into 2 factors one for the Japanese 10-year yield acting as a single source of risk and the other one gathering the UK, US, and German bonds. The stock market factor is separated according to a regional split: factor 3, 5, and 8 represents respectively the global stock market, the Asian and the European regions. 

The factors in dataset 2 are interpreted similarly. In March 2000, the 5 factors represented in Figure \ref{fig:exloadingsbeforecrisis2} are long-only portfolios consisting of US bonds, global stocks indices, US stock indices, commodities, and French bonds, respectively, while in January 2008, we find only two risk factors: one representing the bond market and the other one mostly the stock market.

\begin{figure}[ht!]
	\centering
	\begin{subfigure}[b]{0.49\textwidth}
		\centering
		\includegraphics[width=\linewidth]{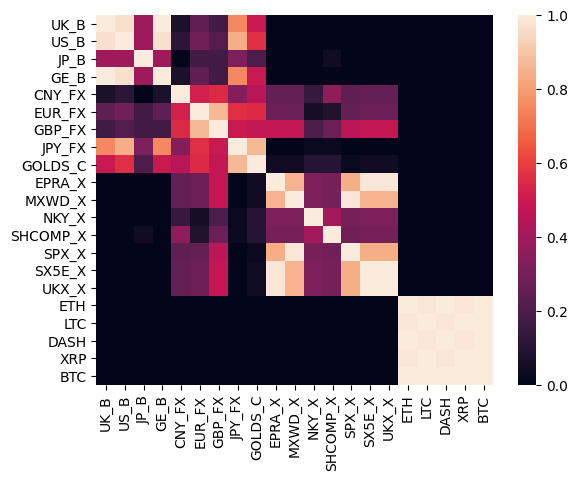}
		\caption{}
		\label{fig:clusterfreq1}
	\end{subfigure}
	\hfill
	\begin{subfigure}[b]{0.49\textwidth}
		\centering
		\includegraphics[width=\linewidth]{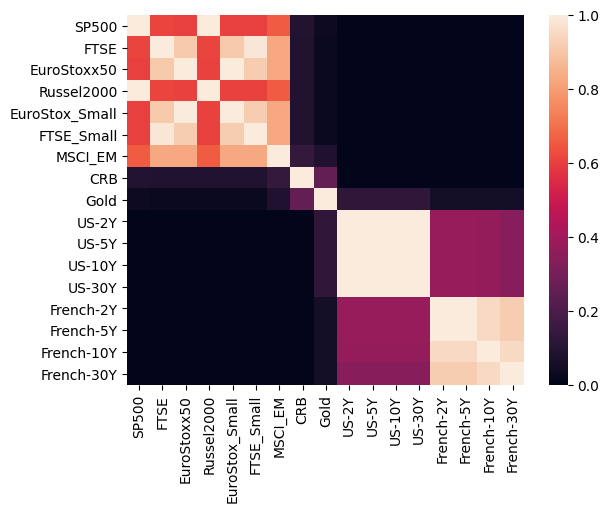}
		\caption{}
		\label{fig:clusterfreq2}
	\end{subfigure}
	\caption[]{Average cluster assignment counts, $M(i,j) = 1/T \sum_{t=1}^T \mathbf{I}\{(i,j) \in \mathcal{C}_k^t,\ 1 \leq k \leq p \}$ where $1\leq i,j \leq d$ and $\mathcal{C}_k = \{i\}_{1 \leq i \leq d,\ W_{ik} > 0}$ \quantletNMFRB}
	\label{fig:cluster_freq}
\end{figure}

The stability of the constitution of the NMF portfolios is represented in Figure \ref{fig:cluster_freq}. We count how many times a pair of assets $(i,j)$ is explained by one factor and average it over the rebalancing periods. For dataset 1 in Figure \ref{fig:clusterfreq1}, we observe that the cryptocurrencies are almost all the time represented by a singular factor, representing the cryptocurrency market. The same observation is made for the US treasury yields in dataset 2 represented in Figure \ref{fig:clusterfreq2}. Similarly, we can identify some groups of assets that are almost never split into the NMF factors like (UK\_B, US\_B, GE\_B), (EUR\_FX, GBP\_FX), (JPY\_FX, GOLDS\_C), (EPRA\_X, SX5E\_X, UKX\_X) in dataset 1 or, the french bonds, (SP500, Russel2000), (FTSE, FTSE\_small), (EuroStoxx50, EuroStoxx\_Small) in dataset 2.

\subsubsection{NMF factors risk contributions}

\begin{figure}[ht!]
	\centering
	\begin{subfigure}[t]{0.49\textwidth}
		\centering
		\includegraphics[width=\linewidth]{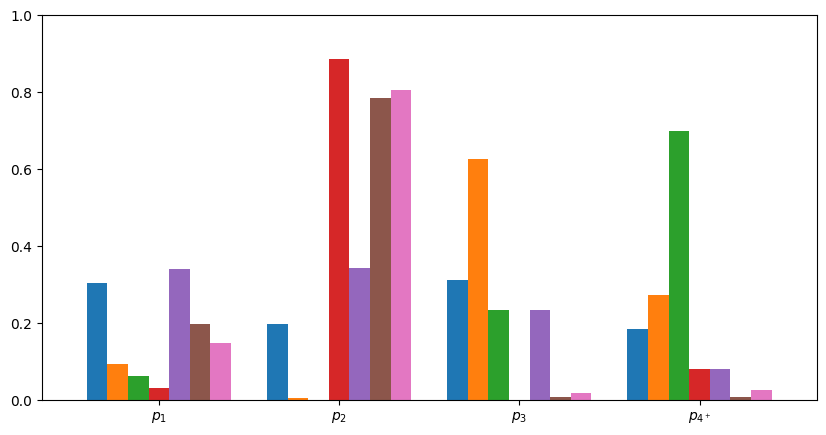}
		\caption{During the COVID crisis in May 2020 in universe 1}
		\label{fig:rcscovid}
	\end{subfigure}
	\hfill
	\begin{subfigure}[t]{0.49\textwidth}
		\centering
		\includegraphics[width=\linewidth]{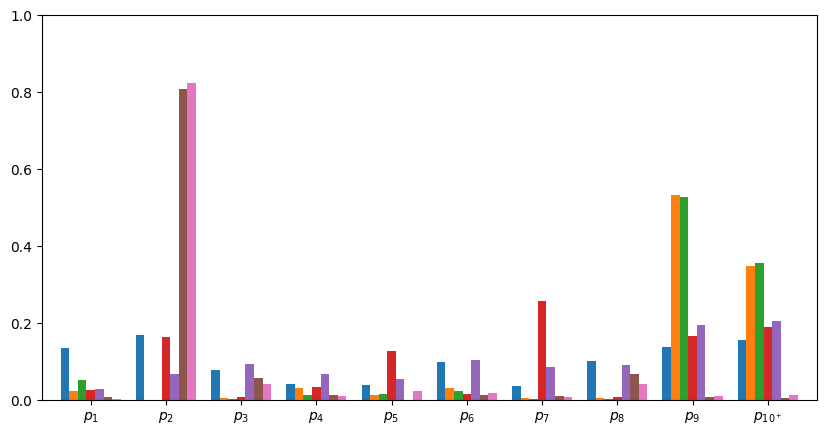}
		\caption{Before Fed's rate increase in December 2021 in universe 1}
		\label{fig:rcsendlowrate2022}
	\end{subfigure}
	\vskip\baselineskip
	
	\begin{subfigure}[t]{0.49\textwidth}
		\centering
		\includegraphics[width=\linewidth]{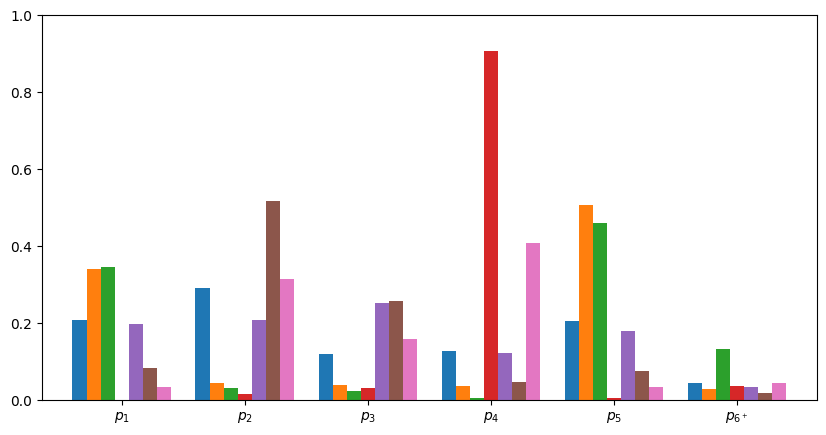}
		\caption{Before the dotcom bubble crash in March 2000 in universe 2}
		\label{fig:rcs_before_com_crash_2}
	\end{subfigure}
	\hfill
	\begin{subfigure}[t]{0.49\textwidth}
		\centering
		\includegraphics[width=\linewidth]{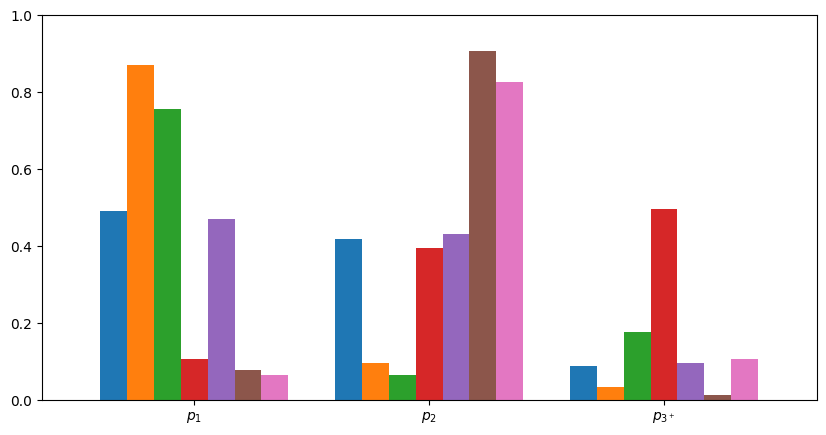}
		\caption{During 2008 crisis in January 2008 in universe 2}
		\label{fig:rcs_during_2008_crisis_2}
	\end{subfigure}
	\caption[]{Risk contribution of the NMF factors for the \textcolor{Blue}{ERC}, \textcolor{Orange}{NMRP}, \textcolor{Green}{HRP}, \textcolor{Red}{HCAA}, \textcolor{Purple}{NMFRB}, \textcolor{Brown}{Equal}, \textcolor{Rhodamine}{Equal Class} portfolios defined in Equation \eqref{eq:rc_nmf}. For simplicity, we represent the risk contribution of the idiosyncratic factors with $m_p^+ = \sum_{k=p+1}^d m_k$ \quantletNMFRB}
	\label{fig:ex_rcs}
\end{figure}

Now that we know how to interpret each factor, let us look at their risk exposure. The graphs in Figure \ref{fig:ex_rcs} represent the risk contributions of the NMF factors (see Equation \eqref{eq:rc_nmf}) for the various strategies with allocations corresponding to the examples in the previous Figure \ref{fig:ex_oos_loadings}. The naive Equal and Equal class strategies, though there are by definition balanced in the portfolio weights, are not balanced in terms of risk contributions. They concentrate the risk in 1 factor out of 3, 1 out of 9, 2 out of 4, and 1 out of 2 in Figure \ref{fig:rcscovid}, \ref{fig:rcsendlowrate2022}, \ref{fig:rcs_before_com_crash_2} and \ref{fig:rcs_during_2008_crisis_2}, accounting for around 80, 80, 70 and 80\% of the portfolio risk, respectively. The factors concentrating the risks are mostly exposed to the crypto and stock markets in dataset 1 and 2, respectively. This behavior is also observed for the HCAA portfolio during the COVID crisis in Figure \ref{fig:rcscovid} where factor 2, which is exposed to the crypto market accounts for around 85\% of the portfolio risk. HRP concentrates the risk on the additional idiosyncratic risk factors in Figures \eqref{fig:rcscovid} and \eqref{fig:rcsendlowrate2022} which account for approximately 70 and 30\% of the portfolio risk, respectively. In comparison, NMFRP allocates more risk to the first common NMF factors. Finally, the NMFRB and ERC portfolios seem to be well-balanced. 

\begin{figure}[ht!]
	\centering
	\begin{subfigure}[t]{0.49\textwidth}
		\centering
		\includegraphics[width=\linewidth]{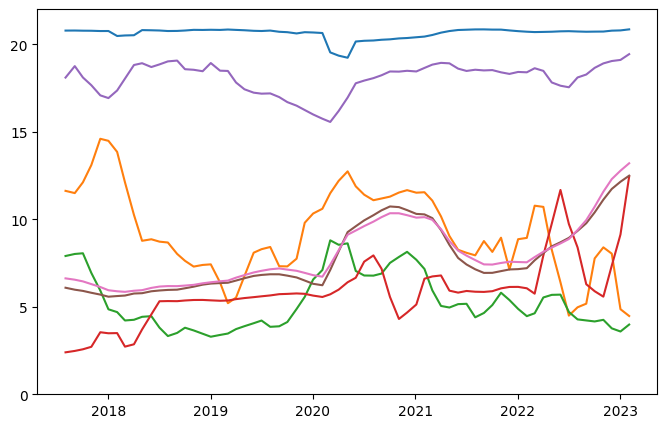}
		\caption{MT Bets dataset 1}
		\label{fig:mintorsionain1}
	\end{subfigure}
	\hfill
	\begin{subfigure}[t]{0.49\textwidth}
		\centering
		\includegraphics[width=\linewidth]{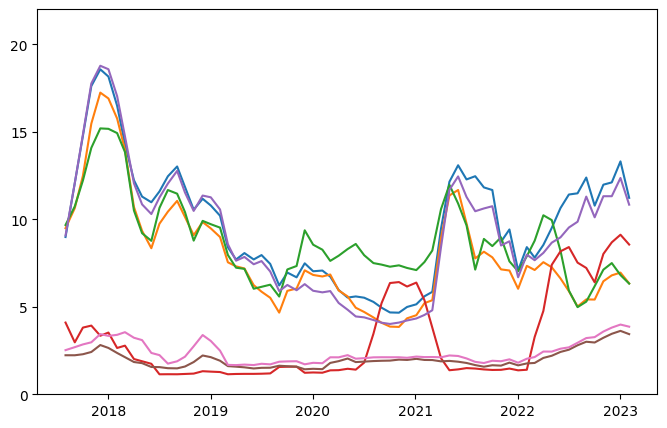}
		\caption{$\operatorname{MT}_F$ Bets dataset 1}
		\label{fig:mintorsionfin1}
	\end{subfigure}
	\vskip\baselineskip
	
	\begin{subfigure}[t]{0.49\textwidth}
		\centering
		\includegraphics[width=\linewidth]{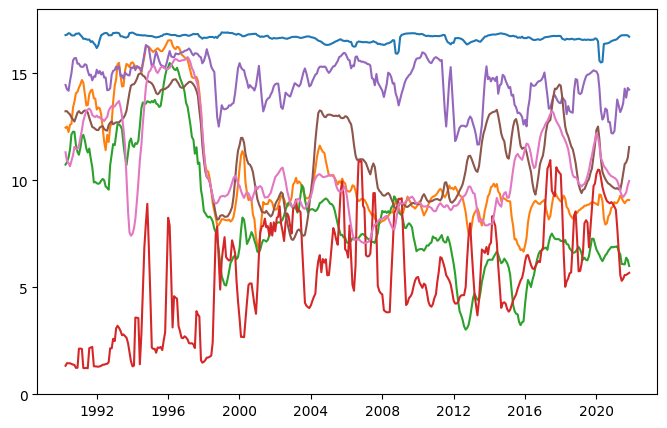}
		\caption{MT Bets dataset 2}
		\label{fig:mintorsionain2}
	\end{subfigure}
	\hfill
	\begin{subfigure}[t]{0.49\textwidth}
		\centering
		\includegraphics[width=\linewidth]{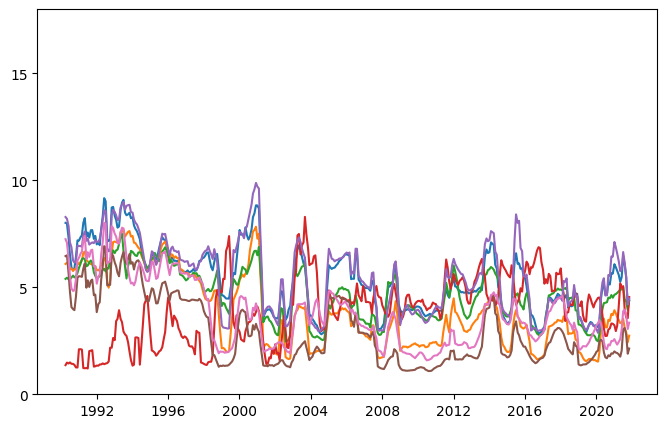}
		\caption{$\operatorname{MT}_F$ Bets dataset 2}
		\label{fig:mintorsionfin2}
	\end{subfigure}
	\caption[]{3 months rolling average of MT (left) and $\operatorname{MT}_F$ (right) Bets for dataset 1 (top) and dataset 2 (bottom) for the \textcolor{Blue}{ERC}, \textcolor{Orange}{NMRP}, \textcolor{Green}{HRP}, \textcolor{Red}{HCAA}, \textcolor{Purple}{NMFRB}, \textcolor{Brown}{Equal}, \textcolor{Rhodamine}{Equal Class} portfolios \quantletNMFRB}
	\label{fig:bets_ts}
\end{figure}

Finally, we study the diversification degree in uncorrelated MT and $\operatorname{MT_F}$ Bets in Figure \ref{fig:bets_ts}. We observe on both datasets that MT Bets is relatively stable and close to the upper bound for the ERC and NMFRB portfolios, while HRP, NMFRP, and HCAA MT Bets are more volatile. Also, we notice on both datasets that, at the asset level, HCAA is not diversified at the beginning of the periods and that the diversification score of inverse-variance allocations (HRP and NMFRP) seems to exhibit a downward trend. This is strong evidence that NMFRB and ERC are diversified portfolios at the asset level for across various market regimes. However, it seems that the inter-factor diversification score is affected by market regimes: we observe on dataset 1 that $\operatorname{MT\ Bets}_F(\operatorname{NMFRB})$ attains its minimum during the COVID crisis in 2020. Similarly, we observe a sharp decline of the $\operatorname{MT\ Bets}_F(\operatorname{NMFRB})$ after the "dotcom" crash in the early 2000s. 

%% file: mc_sim_nmf.tex
\section{Robustness check}
\subsection{Out-of-sample Monte Carlo simulations}

In this section, we investigate the robustness of the proposed portfolio allocation method further. As it is too easy to chose a historical sample where our strategy would perform best \cite{BaileyPrado:2015}, we simulate multiple historical paths via Monte Carlo simulation. In order to fairly compare our method with the HRP proposed by \cite{Prado:2016}, we use a similar simulation as the one described in López de Prado's paper.

First, we generate 20 time series of 520 observations with different volatility levels forming 5 clusters with high pairwise correlation within each cluster and low correlation between clusters. To do so we fit a $\operatorname{GARCH}(1,1)$ process on the log-returns of $\mathcal{S} = \{\operatorname{BTC}, \operatorname{SPX\_X}, \operatorname{US\_B}, \operatorname{EUR\_FX}, \operatorname{GOLD\_C}\}$ from dataset 1 and simulate new observations $\{x_i\}_{i \in \mathcal{S}}$ with the estimated $\operatorname{GARCH}$ parameters. For each, $x_i$, we generate $n_i$ time series with $\widetilde{x}_i = x_i + \mathcal{N}(0, \sigma_{\widetilde{x}_i}^2)$ and $\sigma_{\widetilde{x}_i} \simeq \sigma_{x_i}$ forming in total 15 time series that are clustered with the original one. Finally, as in López de Prado's paper, we add random common and idiosyncratic shocks to the generated returns.

Then, we compute $\operatorname{HRP}$, $\operatorname{HCAA}$, $\operatorname{NMFRB}$ portfolios by looking back at 260 observations. These portfolios are re-estimated and rebalanced monthly (every 22 observations). 

We compute the out-of-sample returns associated with the portfolios with a $5\%$ annualized volatility target, thanks to Equation \eqref{eq:levered_weights}. We add a constraint on the leverage factor setting a maximum factor of 3. This procedure is repeated 10000 times.

\begin{figure}
		\centering
		\begin{subfigure}[b]{0.24\textwidth}
			\centering
			\includegraphics[width=\textwidth]{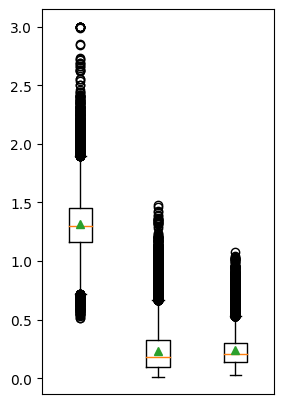}
			\caption[leverage]%
			{Leverage factor}    
			\label{fig:mc leverage}
		\end{subfigure}
		\hfill
		\begin{subfigure}[b]{0.24\textwidth}
			\centering
			\includegraphics[width=\textwidth]{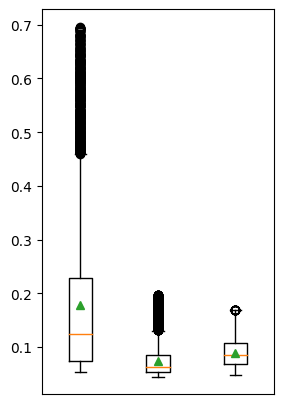}
			\caption[volatility]%
			{Annualized Volatility}    
			\label{fig:mc volatility}
		\end{subfigure}
		\hfill
		\begin{subfigure}[b]{0.24\textwidth}  
			\centering 
			\includegraphics[width=\textwidth]{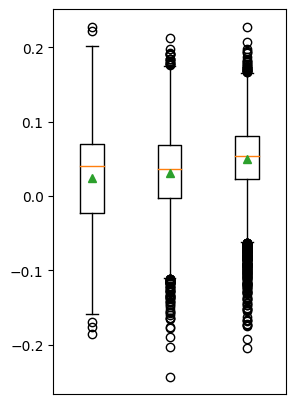}
			\caption[Adjusted Sharpe ratio]%
			{Adjusted Sharpe ratio}    
			\label{fig:mc Adjusted Sharpe ratio}
		\end{subfigure}
		\hfill
		\begin{subfigure}[b]{0.24\textwidth}   
			\centering 
			\includegraphics[width=\textwidth]{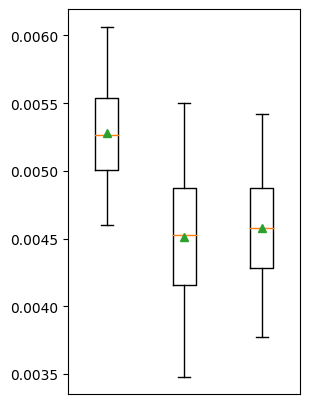}
			\caption[]%
			{ $\operatorname{VaR}(5\%)$}    
			\label{fig:mc var}
		\end{subfigure}
		\vskip\baselineskip
		\begin{subfigure}[b]{0.24\textwidth}   
			\centering 
			\includegraphics[width=\textwidth]{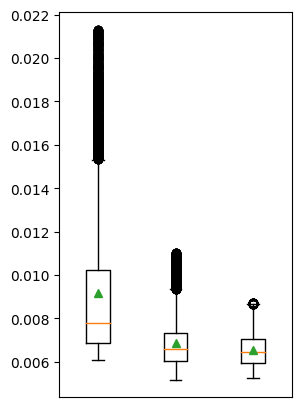}
			\caption[]%
			{$\operatorname{ES}(5\%)$}    
			\label{fig:mc es}
		\end{subfigure}
		\hfill
		\begin{subfigure}[b]{0.24\textwidth}
			\centering
			\includegraphics[width=\textwidth]{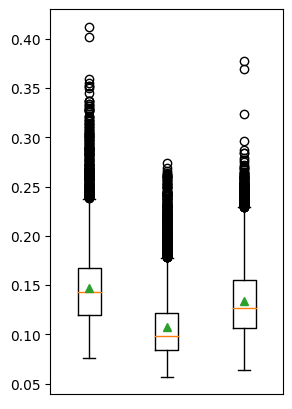}
			\caption[total returns]%
			{$\operatorname{SSPW}$}    
			\label{fig:mc sspw}
		\end{subfigure}
		\hfill
		\begin{subfigure}[b]{0.24\textwidth}   
			\centering 
			\includegraphics[width=\textwidth]{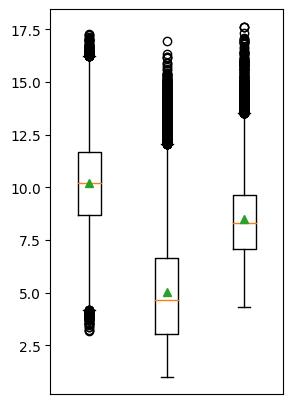}
			\caption[]%
			{$\operatorname{MT\ Bets}$}    
			\label{fig:mc mt_bets}
		\end{subfigure}
		\hfill
		\begin{subfigure}[b]{0.24\textwidth}   
		\centering 
		\includegraphics[width=\textwidth]{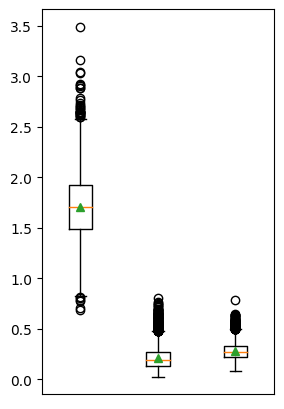}
		\caption[]%
		{$\operatorname{TTO}$}    
		\label{fig:mc tato}
		\end{subfigure}
	
		\caption[]{Metrics over Monte Carlo samples for $\operatorname{HRP}$, $\operatorname{HCAA}$, $\operatorname{NMFRB}$ from left to right. The boxplots are computed on the observations lying between the 5 and 95\% percentiles for a better readability, the \textcolor{Green}{mean} is represented in \textcolor{Green}{green} and the \textcolor{Orange}{median} in \textcolor{Orange}{orange} \quantletNMFRB}
		\label{fig:montecarlo}
\end{figure}

Figure \ref{fig:montecarlo} illustrates that the $\operatorname{NMFRB}$ portfolio with a 5\% target volatility outperforms $\operatorname{HRP}$ in terms of risk-adjusted returns and robustness. Figures \ref{fig:mc leverage} and \ref{fig:mc volatility} demonstrate that the $\operatorname{NMFRB}$ strategy has a more stable distribution of returns, with an average annualized volatility below 10\%, whereas the $\operatorname{HRP}$ has a less stable volatility, averaging at 18\% across the Monte Carlo simulations. Although all three strategies struggle to reach the 5\% volatility target, it is more difficult for $\operatorname{HRP}$, which requires much larger leverage, as presented in Figure \ref{fig:mc leverage}, due to its lower inherent volatility compared to the HCAA and NMFRB strategies without volatility target. Furthermore, $\operatorname{NMFRB}$ demonstrates a better risk-adjusted profile with a more stable and larger average Sharpe ratio. Finally, as shown in Figures \ref{fig:mc var} and \ref{fig:mc es}, the tail risk of $\operatorname{NMFRB}$ is lower than that of $\operatorname{HRP}$: on both figures the inter-quartile ranges of $\operatorname{VaR}(5\%)$ and $\operatorname{ES}(5\%)$ do not overlap, with an average $\operatorname{VaR}(5\%)$ of $\operatorname{NMFRB}$ being 0.46\%, which is 15\% lower than that of $\operatorname{HRP}$, and an average $\operatorname{ES}(5\%)$ of $\operatorname{NMFRB}$ being 55\% lower than $\operatorname{HRP}$. As shown in Figure \ref{fig:mc tato}, even when considering transaction costs, the outperformance of $\operatorname{NMFRB}$ over $\operatorname{HRP}$ still holds, as its $\operatorname{TTO}$ is approximately five times that of $\operatorname{HRP}$. In this simulation, HCAA and NMFRB demonstrate comparable performance in terms of risk-adjusted returns. 

Finally, HRP seems slightly more diversified than NMFRB, with an average of 10.1 and 8.5 MT bets, respectively, and a similar variance. The SSPW diversification is similar for both (0.15 and 0.13, respectively). However, HCAA is much less diversified in terms of effective bets, making only half as many as HRP.

In summary, in this simulation, NMFRB outperforms HRP in several ways. NMFRB requires less leverage, maintains more stable weights, achieves better risk-adjusted returns, and has lower tail risk. Additionally, NMFRB is nearly as well-diversified as HRP which indicates that it strikes a balance between diversification across all assets and diversification across clusters of investments. These empirical results suggest that NMFRB may offer superior protection against both common and idiosyncratic shocks.

%% file: conclusion_nmf.tex
\section{Conclusion}

In this paper, we build a new long-only portfolio allocation method based on a dynamic sparse linear factor model for the returns via non-negative matrix factorization. The factors can easily be interpreted as long-only portfolio or synthetic asset classes that cluster highly correlated assets. By making simple and reasonable assumptions on the discovered correlation structure, we propose a new portfolio allocation, called NMF Risk Budget (NMFRB), that allocates equal risk between the synthetic asset classes and more risk to the assets that are the better represented. Thanks to this diversification approach, we obtain an allocation which is diversified at both the latent factors and original assets levels. We show that the factors are stable in their interpretation and bring out persistent correlations between assets. On top, they robustly provide effective diversification across various market regimes even if they are not perfectly uncorrelated. This method outperforms HRP in terms of risk-adjusted returns on a long history.

%% file: appendix_nmf.tex
\appendix

\section{Portfolio weights}
\begin{figure}[ht!]
	\centering
	\begin{subfigure}[b]{0.32\textwidth}
		\centering
		\includegraphics[width=\textwidth]{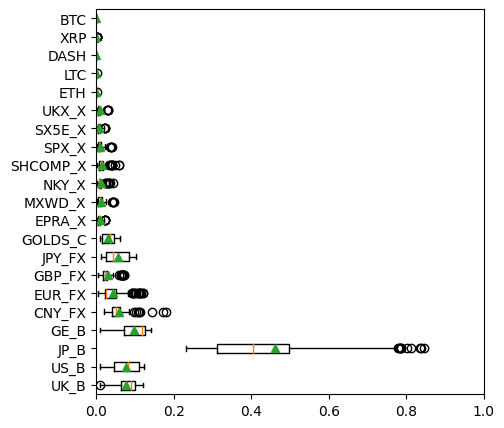}
		\caption{NMFRP}
		\label{fig:weightsnmfrp_1}
	\end{subfigure}
	\hfill
	\begin{subfigure}[b]{0.32\textwidth}
		\centering
		\includegraphics[width=\textwidth]{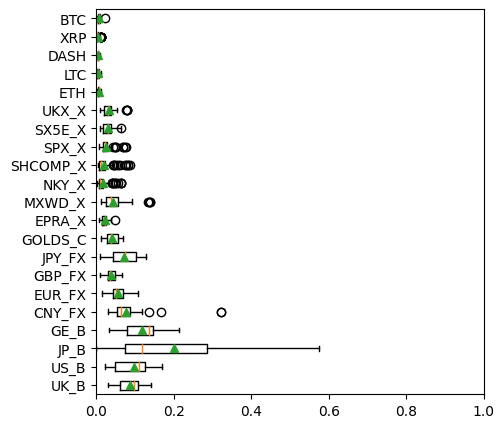}
		\caption{NMFRB}
		\label{fig:weightsrbfactor_1}
	\end{subfigure}	
	\begin{subfigure}[b]{0.32\textwidth}
		\centering
		\includegraphics[width=\textwidth]{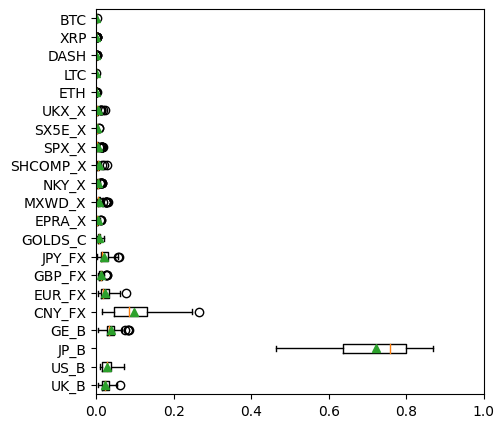}
		\caption{HRP}
		\label{fig:weightshrp_1}
	\end{subfigure}
	\vskip\baselineskip

	\begin{subfigure}[b]{0.32\textwidth}
		\centering
		\includegraphics[width=\textwidth]{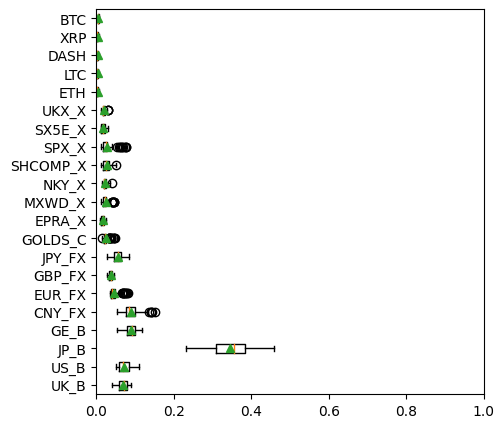}
		\caption{ERC}
		\label{fig:weightserc_1}
	\end{subfigure}
	\hfill
	\begin{subfigure}[b]{0.32\textwidth}
		\centering
		\includegraphics[width=\textwidth]{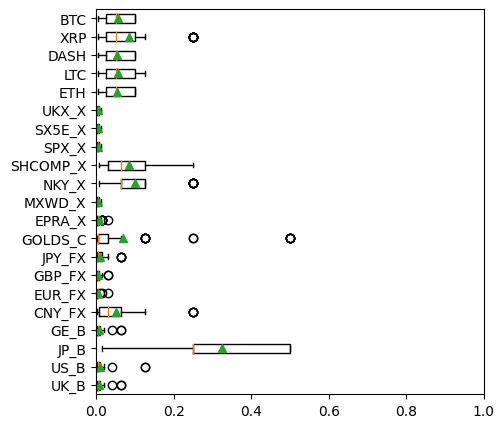}
		\caption{HCAA}
		\label{fig:weightshcaa_1}
	\end{subfigure}
	\hfill
	\begin{subfigure}[b]{0.32\textwidth}
	\centering
	\includegraphics[width=\textwidth]{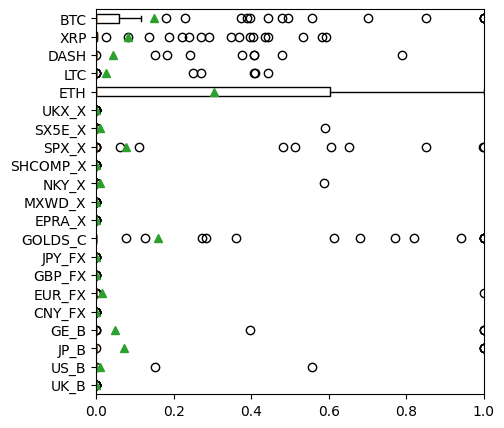}
	\caption{Markowitz}
	\label{fig:weightsmarkowitz_1}
	\end{subfigure}
	\vskip\baselineskip
	
	\begin{subfigure}[b]{0.32\textwidth}
		\centering
		\includegraphics[width=\textwidth]{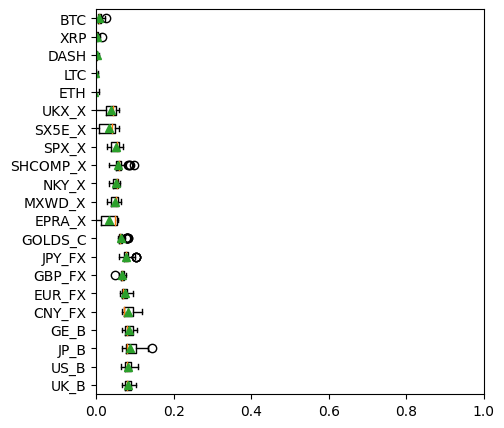}
		\caption{Robust-M}
		\label{fig:weightsgmv_robust_1}
	\end{subfigure}
	\caption[]{Assets weights of the portfolios on dataset 1 \quantletNMFRB}
	\label{fig:weights_1}
\end{figure}

\begin{figure}[ht!]
	\centering
	\begin{subfigure}[b]{0.32\textwidth}
		\centering
		\includegraphics[width=\textwidth]{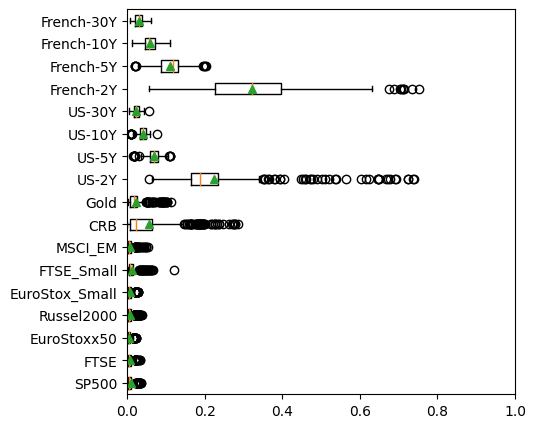}
		\caption{NMFRP}
		\label{fig:weightsnmfrp_2}
	\end{subfigure}
	\hfill
	\begin{subfigure}[b]{0.32\textwidth}
		\centering
		\includegraphics[width=\textwidth]{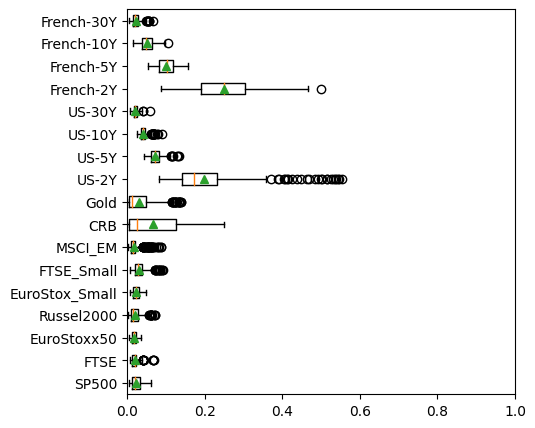}
		\caption{NMFRB}
		\label{fig:weightsrbfactor_2}
	\end{subfigure}	
	\begin{subfigure}[b]{0.32\textwidth}
		\centering
		\includegraphics[width=\textwidth]{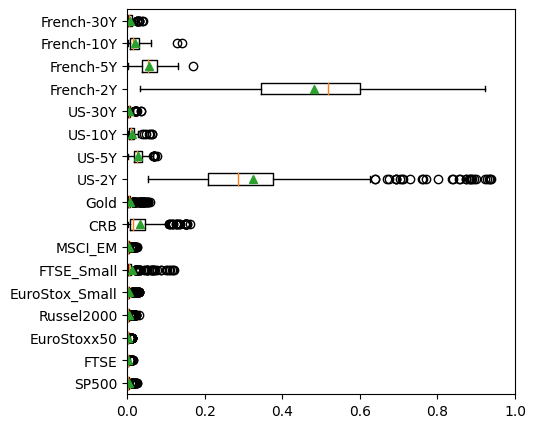}
		\caption{HRP}
		\label{fig:weightshrp_2}
	\end{subfigure}
	\vskip\baselineskip
	
	\begin{subfigure}[b]{0.32\textwidth}
		\centering
		\includegraphics[width=\textwidth]{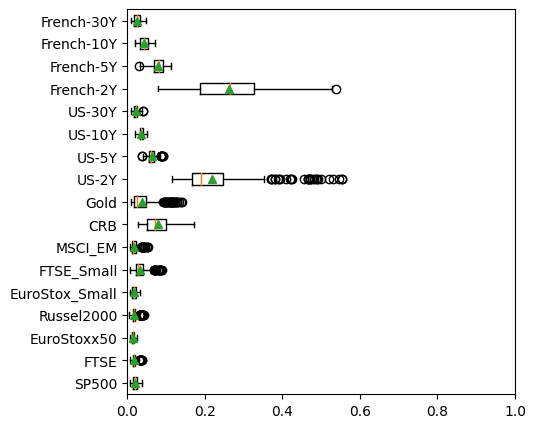}
		\caption{ERC}
		\label{fig:weightserc_2}
	\end{subfigure}
	\hfill
	\begin{subfigure}[b]{0.32\textwidth}
		\centering
		\includegraphics[width=\textwidth]{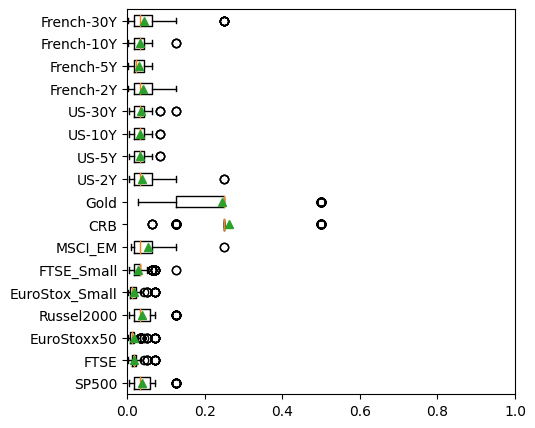}
		\caption{HCAA}
		\label{fig:weightshcaa_2}
	\end{subfigure}
	\hfill
	\begin{subfigure}[b]{0.32\textwidth}
		\centering
		\includegraphics[width=\textwidth]{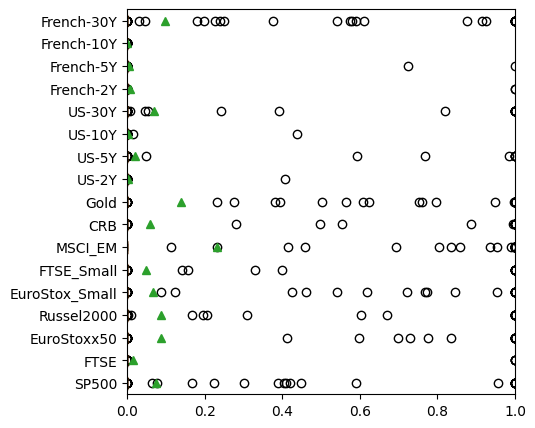}
		\caption{Markowitz}
		\label{fig:weightsmarkowitz_2}
	\end{subfigure}
	\vskip\baselineskip
	
	\begin{subfigure}[b]{0.32\textwidth}
		\centering
		\includegraphics[width=\textwidth]{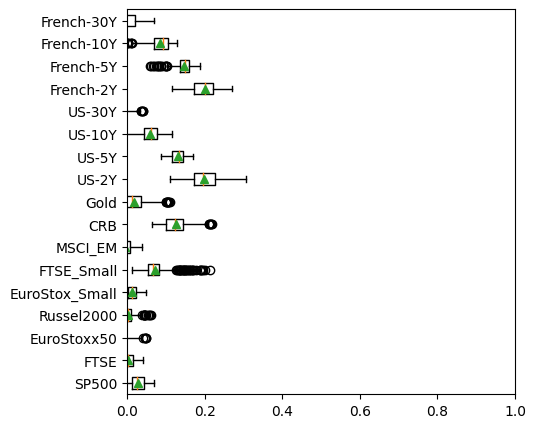}
		\caption{Robust-M}
		\label{fig:weightsgmv_robust_2}
	\end{subfigure}
	\caption[]{Assets weights of the portfolios on dataset 2 \quantletNMFRB}
	\label{fig:weights_2}
\end{figure}